\documentclass[useAMS,usenatbib]{mn2e}

\usepackage{graphicx}

\newcommand{\eq}[1]{\begin{equation}  #1 \end{equation}}

\newcommand{\eqa}[1]{\begin{eqnarray}   #1 \end{eqnarray}}

\newcommand{\br}[1]{\left( #1 \right)}
\newcommand{\bc}[1]{\left\{ #1 \right\}}
\newcommand{\bb}[1]{\left[ #1 \right]}
\newcommand{\ba}[1]{\left\langle #1 \right\rangle}

\newcommand{\nn}{\nonumber}

\newcommand{\dd}{{\rm d}}
\newcommand{\expo}[1]{~{\rm e}^{ #1 }}
\newcommand{\vek}[1]{\mbox{\boldmath $#1$}}
\newcommand{\svek}[1]{\mbox{\boldmath \scriptsize $#1$}}  

\title[Modelling intrinsic alignment contamination of weak lensing surveys]{Intrinsic galaxy shapes and alignments II: Modelling the intrinsic alignment contamination of weak lensing surveys}
\author[B. Joachimi et al.]{B.~Joachimi,$^1$\thanks{E-mail: bj@roe.ac.uk} E.~Semboloni,$^2$ S.~Hilbert,$^{3,4}$ P.E.~Bett,$^5$ J.~Hartlap,$^5$ H.~Hoekstra,$^2$ \newauthor{and P.~Schneider$^5$}\\
  $^1$Institute for Astronomy, University of Edinburgh, Royal Observatory, Blackford Hill, Edinburgh, EH9 3HJ, U.K.\\
  $^2$Leiden Observatory, Leiden University, P.O. Box 9513, 2300 RA, The Netherlands\\
  $^3$Kavli Institute of Particle Astrophysics and Cosmology (KIPAC), Stanford University, 452 Lomita Mall, Stanford, CA 94305, \&\\ SLAC National Accelerator Laboratory, 2575 Sand Hill Road, M/S 29, Menlo Park, CA 94025\\
  $^4$Max-Planck-Institut f{\"u}r Astrophysik, Karl-Schwarzschild-Stra{\ss}e 1, 85741 Garching, Germany\\
  $^5$Argelander-Institut f\"ur Astronomie, Universit\"at Bonn, Auf dem H\"ugel 71, 53121 Bonn, Germany
}
\date{Accepted . Received ; in original form }

\pagerange{\pageref{firstpage}--\pageref{lastpage}} \pubyear{2013}

\begin{document}
\label{firstpage}

\maketitle

\begin{abstract}
Intrinsic galaxy alignments constitute the major astrophysical systematic of forthcoming weak gravitational lensing surveys but also yield unique insights into galaxy formation and evolution. We build analytic models for the distribution of galaxy shapes based on halo properties extracted from the Millennium Simulation, differentiating between early- and late-type galaxies as well as central galaxies and satellites. The resulting ellipticity correlations are investigated for their physical properties and compared to a suite of current observations. The best-faring model is then used to predict the intrinsic alignment contamination of planned weak lensing surveys. We find that late-type galaxy models generally have weak intrinsic ellipticity correlations, marginally increasing towards smaller galaxy separation and higher redshift. The signal for early-type models at fixed halo mass strongly increases by three orders of magnitude over two decades in galaxy separation, and by one order of magnitude from $z=0$ to $z=2$. The intrinsic alignment strength also depends strongly on halo mass, but not on galaxy luminosity at fixed mass, or galaxy number density in the environment. We identify models that are in good agreement with all observational data, except that all models over-predict alignments of faint early-type galaxies. The best model yields an intrinsic alignment contamination of a \textit{Euclid}-like survey between $0.5-10\,\%$ at $z > 0.6$ and on angular scales larger than a few arcminutes. Cutting $20\,\%$ of red foreground galaxies using observer-frame colours can suppress this contamination by up to a factor of two.
\end{abstract}


\section{Introduction}
\label{sec:intro}

Cosmic shear, the weak gravitational lensing effect by the large-scale structure of the Universe, will be one of the primary probes in most of the forthcoming cosmological galaxy surveys, including KiDS\footnote{\texttt{http://www.astro-wise.org/projects/KIDS}}, DES\footnote{\texttt{http://www.darkenergysurvey.org}}, HSC\footnote{\texttt{http://www.naoj.org/Projects/HSC/index.html}}, LSST\footnote{\texttt{http://www.lsst.org}}, and \textit{Euclid}\footnote{\texttt{http://www.euclid-ec.org}} \citep{laureijs11}. The correlations of the distortions of distant galaxy images induced by the continuous deflection of their light by the matter distribution along the line of sight are sensitive to both the evolution of structure and the geometry of the cosmos, so that cosmic shear can provide excellent constraints on dark matter, dark energy, and potential deviations from general relativity \citep{albrecht06,peacock06}.

Precision measurements of cosmological parameters need to be matched by an equally superb control of systematic errors. In particular, weak lensing surveys will have to undertake major efforts to control their dominant astrophysical contamination, the intrinsic alignment of galaxy shapes, which mimics the correlations induced by gravitational lensing.

To date, the knowledge about intrinsic alignments is limited, and virtually non-existent at the typical redshifts of galaxy samples used for weak lensing analyses around $z \sim 1$, and at small, non-linear scales from where the majority of cosmological constraints originates \citep[e.g.][]{takada04}. Hence several methods have been designed to remove or calibrate the intrinsic alignment signal with minimal assumptions about the form of the intrinsic correlations \citep[see e.g.][]{king02,king03,bridle07,joachimi08b,joachimi09,bernstein08,joachimi10,zhang10}. 

While most of these approaches constitute valuable consistency checks, it is likely that none is capable of controlling the intrinsic alignment contamination without a significant loss of statistical power of the survey, unless prior constraints on the systematic are available. Therefore a robust and comprehensive model for intrinsic galaxy alignments is paramount to allow for the optimal exploitation of the future rich weak lensing data sets. Even the design of surveys would profit from an improved understanding of intrinsic alignments which, for instance, are the main driver for the requirements on the accuracy of galaxy redshift measurements \citep{laureijs11}.

Moreover, intrinsic alignments potentially constitute an interesting astrophysical signal that probes the dependence of galaxy formation and evolution on environment, and that is complementary to established techniques. This calls for an intrinsic alignment model that can explain e.g. their dependence on redshift, environment, merger history, and physical properties of galaxies.

The \lq classical\rq\ picture of how galaxies acquire certain shapes assumes a dichotomy between spiral and elliptical galaxies. The shapes of the former are determined by their disc whose rotation axis is aligned with the halo angular momentum, which in turn was created via tidal torquing during the assembly of the galaxy \citep{peebles69}. As the resulting angular momentum scales with the square of the quadrupole of the gravitational potential, one expects the correlations between the angular momenta and hence the projected shapes of spiral galaxies to vanish to first order \citep{hirata04}.

The shapes of elliptical galaxies are assumed to follow the generally triaxial shape of their haloes which are subjected to tidal stretching by the surrounding large-scale matter distribution. The projected ellipticity of the galaxy is then proportional to the quadrupole of the gravitational potential, leading to the linear alignment model for early-type galaxies \citep{croft00,heavens00,catelan01}.

Even if these simple prescriptions prove to be valid for correlations at large galaxy separation, one expects modifications on small scales where non-linear structure evolution, baryonic physics, and dynamical processes in high-density regions become relevant. One way forward is to incorporate the alignments of satellite galaxies in a halo model \citep{schneiderm09}, which predicts similar correlations as an earlier empirical ansatz by \citet{bridle07} to employ the full non-linear power spectrum in the linear alignment model.

The predictions of intrinsic alignments by the tidal torque and tidal stretching paradigms are consistent with current observations, which however are still fraught with relatively large uncertainties. Measurements among late-type galaxies in the SDSS Main samples \citep{hirata07} and the overlap region between the SDSS and WiggleZ surveys \citep{mandelbaum11} have not yielded any significant detection, except at very low redshift \citep{lee11}. 

In contrast, correlations between intrinsic shapes as well as the alignment of intrinsic shapes towards overdensities of early-type galaxies have been robustly detected in several data sets out to redshift 0.7 (\citealp{brown02,heymans04,mandelbaum06,hirata07,okumura08,okumura09,joachimi11,li13}; see also \citealp{heymans13} for a detection in cosmic shear data). The observed correlation functions are in agreement with the galaxy separation and redshift dependencies predicted by the linear alignment model and its non-linear modifications (\citealp{joachimi11}; see also \citealp{blazek11}), featuring in addition a clear increase in amplitude with galaxy luminosity \citep{hirata07,joachimi11}.

Concerning small scales, observations of satellite galaxy shape alignment towards the centre of clusters reach back to \citet{hawley75}. Recent results range between weakly significant radial alignment of satellites \citep{pereira05,agustsson06,faltenbacher07,hao11} and non-detection of anisotropy in the orientations \citep{bernstein02b,hung11}, with uncertain levels of selection effects and systematic errors.

A successful model of intrinsic galaxy shapes has to fit large-scale correlations of galaxy ellipticities, small-scale satellite alignment, as well as the distribution of galaxy ellipticities simultaneously. In \citet{joachimi12}, Paper I hereafter, dark matter halo properties from the Millennium Simulation were combined with semi-analytic models of galaxy evolution and analytic models for the link between the shape of galaxies and their underlying halo. The resulting shape models were confronted with one-point statistics and distributions of galaxy ellipticities measured in the COSMOS Survey.

In this work we continue to follow this ansatz of \lq semi-analytic\rq\ galaxy shape modelling, now investigating the two-point statistics of ellipticities. The Millennium Simulation \citep{springel05} is well suited to build our galaxy shape models because it has a sufficiently large volume to allow for measurements of correlations among widely separated galaxies, combined with good mass resolution for accurate halo shape and angular momentum measurements (cf. \citealp{heymans06} who use a similar approach but rely on simulations that have 20 times higher particle mass).

Numerous publications \citep[e.g.][]{bailin05,altay06,hahn07b,lee08} have investigated the correlations of dark matter halo ellipticities and angular momenta with each other, and with the large-scale matter distribution, in N-body simulations. To enable an accurate selection of galaxy samples for comparison with observations, we supplement this information with multi-band photometry and galaxy type classifications from the semi-analytic models of galaxy formation and evolution by \citet{bower06}.

To incorporate the critical link between the shape of the visible, baryonic matter and the structure of the underlying dark matter distribution, we make use of the statistical properties of this relation extracted from a large number of small-scale, high-resolution hydrodynamic simulations by \citet{bosch02,croft09,hahn10,bett10,bett11}. Furthermore, as the Millennium data includes the positions of satellite galaxies, but not the shapes of the corresponding subhaloes, we resort to simple models of satellite shapes and alignments, based on the high-resolution simulations by \citet{knebe08} whose findings are qualitatively in agreement with similar investigations by \citet{kuhlen07,pereira08,faltenbacher08,knebe10}.

This article is structured as follows. In Section$~$\ref{sec:sims} we briefly summarise the main aspects of the underlying simulations, before providing in Section$~$\ref{sec:galaxymodels} a synopsis of our models of galaxy shapes. Section$~$\ref{sec:simresults} highlights the dependence of the shape correlations resulting from the simulations on redshift, mass, luminosity, and environment. In Section$~$\ref{sec:iaobs} the intrinsic alignments in WiggleZ and SDSS galaxy samples are compared to the simulation-based alignment signals. The impact of the best-matching intrinsic alignment model on weak lensing surveys is studied in Section$~$\ref{sec:impact}. In Section$~$\ref{sec:conclusions} we summarise and conclude on our findings.

Unless stated otherwise, rest-frame magnitudes are $k+e$-corrected to $z=0$ and computed assuming the cosmology of the Millennium Simulation (see below) except a Hubble constant $H_0=100h\, {\rm km/s/Mpc}$ with $h=1$. Magnitudes extracted from the Millennium database are given in the Vega system, while all observations use the AB system. If direct comparison is necessary, we resort to the conversion tables of \citet{fukugita96}. When multiple data sets with error bars are plotted, points are slightly offset horizontally for clarity throughout.

\section{Simulations}
\label{sec:sims}

In this section we briefly summarise the key aspects of the simulations that we use. More details are provided in Section$~$2 of Paper I. The basis of our galaxy shape models is the Millennium Simulation \citep{springel05} which provides us both with the volume (comoving box size $500\,{\rm Mpc}/h$) to measure correlations on cosmological scales and with the mass resolution (particle mass $m_{\rm p}=8.6 \times 10^8\,h^{-1} M_\odot$) needed to determine the properties of galaxy-sized dark matter haloes accurately. We will work with the 32 snapshot outputs from $z \approx 2.1$ to $z=0$.

The cosmology of the Millennium Simulation follows a spatially flat $\Lambda$CDM model with matter density parameter $\Omega_{\rm m}=1-\Omega_\Lambda=0.25$. The baryon density parameter is $\Omega_{\rm b}=0.045$, the Hubble parameter $h=0.73$, the power-law index of the initial power spectrum $n_{\rm s}=1$, and the normalisation of the power of matter fluctuations $\sigma_8=0.9$. This latter value is high compared to recent observational constraints suggesting $\sigma_8 \sim 0.8$ \citep[e.g.][]{planck13-XVI,heymans13}, but the impact of $\sigma_8$ on the strength of intrinsic galaxy correlations is unknown at present (see Paper I for a discussion on the expected effects of high $\sigma_8$ as well as of baryons on intrinsic alignments). We will compare the simulation-based intrinsic alignment signals directly to observations, thereby assuming that the high value of $\sigma_8$ does not affect our results significantly. Note however that, when calculating the importance of intrinsic alignments relative to cosmic shear correlations, we will use $\sigma_8=0.9$ also for the latter to be consistent.

Models of galaxies were placed into the Millennium haloes via the semi-analytic galaxy evolution model \texttt{GALFORM} in the version of \citet{bower06}. From these models we extract apparent and rest-frame magnitudes in various bands as well as the distinction between central and satellite galaxies, where the former are defined as the galaxy in the most massive substructure of a halo at any given time.

Moreover, we adopt the classification of galaxy morphologies via the bulge-to-total ratio of $K$-band luminosity, $R_{\rm type}=L_{K, {\rm bulge}}/L_{K, {\rm total}}$, as proposed by \citet{parry09}. We define our \lq early-type\rq\ galaxy sample via $R_{\rm type}>0.6$ and the \lq late-type\rq\ sample accordingly by $R_{\rm type}<0.6$. The latter comprises spiral and lenticular galaxies, the cut at $R_{\rm type}=0.6$ being motivated by the strikingly different merger histories of these populations \citep[see][]{parry09}.

Ray-tracing is employed to compute the gravitational shear at the positions of dark matter haloes, generated by the light deflection of the matter distribution along the line of sight. The method is detailed in \citet{hilbert09} and provides us with 64 light cones with an area of $4 \times 4\,{\rm deg}^2$ each, yielding a mock survey of $1024\,{\rm deg}^2$ out to a redshift of $z \approx 2.1$. Special care has been taken to avoid that parts of dark matter haloes end up on different lens planes which in the context of this paper could e.g. create artificial alignments of close halo/galaxy pairs. All vectorial quantities such as satellite positions, shape ellipsoid major axes, or angular momenta are transformed into a coordinate system with the line of sight as one basis vector, which facilitates projection onto the plane of the sky.

\section{Galaxy shape modelling}
\label{sec:galaxymodels}

In the following we provide a synopsis of the various shape models which we investigate. Here we will focus on the alignment properties, while the assignment of galaxy shapes has been laid out in detail in Section$~$3 of Paper I. Generally, we divide the galaxy sample into central and satellite galaxies, as identified by the semi-analytic models, and further distinguish between early- and late-type galaxies based on the bulge-to-total luminosity, $R_{\rm type}$. An overview on the model variants used and their naming conventions is given in Table~\ref{tab:galaxymodels}.

\begin{table*}
\centering
\caption{Overview on models for galaxy shapes. The distinction between central and satellite galaxy is adopted from the \texttt{GALFORM} semi-analytic model. Early- and late-type galaxies are distinguished via the bulge-to-total luminosity, $R_{\rm type}$. Shape models for the full galaxy samples are described by combinations of the three-character names listed in the fourth column. Note that low-mass galaxies with too few particles in their haloes to make accurate shape and angular momentum measurements are assigned random orientations, but otherwise follow the model assumed for the respective central galaxy type.}
\begin{tabular}[t]{llll}
\hline\hline
halo type & galaxy type & model & identifier\\
\hline
 & early-type & same shape as halo; simple inertia tensor; galaxy aligned & \texttt{Est}\\
 & " & same shape as halo; reduced inertia tensor; galaxy aligned & \texttt{Ert}\\
 & " & same shape as halo; simple inertia tensor; \citet{okumura08} misalignment & \texttt{Ema}\\
central & late-type & thick disc $\perp$ angular momentum; $r_{\rm edge-on}=0.25$; galaxy aligned & \texttt{Sal}\\
 & " & thick disc $\perp$ angular momentum; $r_{\rm edge-on}=0.25$; \citet{bett11} misalignment & \texttt{Sma}\\
 & " & thick disc $\perp$ angular momentum; $r_{\rm edge-on}=0.1$; \citet{bett11} misalignment & \texttt{Sth}\\
\hline
 & early-type & major axis $\rightarrow$ halo centre; shape sampled from MS halo distribution; simple inertia tensor & \texttt{est}\\
 & " & major axis $\rightarrow$ halo centre; shape sampled from MS halo distribution; reduced inertia tensor & \texttt{ert}\\
satellite & " & major axis $\rightarrow$ halo centre; \citet{knebe08} shape modifications \& misalignment & \texttt{ekn}\\
 & late-type & thick disc pointing to halo centre; $r_{\rm edge-on}=0.25$; galaxy aligned & \texttt{sal}\\
 & " & thick disc pointing to halo centre; $r_{\rm edge-on}=0.25$; \citet{bett11} misalignment & \texttt{sma}\\
 & " & thick disc pointing to halo centre; $r_{\rm edge-on}=0.1$; \citet{bett11} misalignment & \texttt{sth}\\
\hline
\end{tabular}
\label{tab:galaxymodels}
\end{table*}

\subsection{Halo shapes and angular momenta}
\label{sec:shapes}

We base our models of galaxy shapes on the morphologies and angular momenta of the underlying dark matter haloes. The same approach as in \citet{bett07} is used to identify bound structures in the matter distribution of the simulation and to compute the shapes and angular momenta of haloes. A friends-of-friends algorithm is employed to construct groups of simulation particles within which self-bound structures are detected with the \texttt{SUBFIND} code \citep{springel05}. A halo is then defined as a collection of self-bound sub-haloes.

A halo shape is computed via the quadrupole tensor of the mass distribution $\mathbf{M}$ with components
\eq{
\label{eq:massquadrupole}
M_{\mu\nu} = m_{\rm p} \sum_{i=1}^{N_{\rm p}} r_{i,\mu} r_{i,\nu}\;,
}
where $N_{\rm p}$ is the number of all particles that belong to the halo, and where $\vek{r}_i$ denotes the position vector of particle $i$ with respect to the halo centre (defined as the location of the gravitational potential minimum). The eigenvalues and eigenvectors of $\mathbf{M}$ define an ellipsoid, with the eigenvalues per unit mass giving the square of the semi-axis lengths, and the corresponding eigenvectors specifying the axis orientations. We interpret this ellipsoid as an approximation to the shape of the halo.

Our input catalogues contain all haloes in the Millennium database that host a galaxy, with particle numbers down to 20. In Paper I it was demonstrated that a minimum particle number of $N_{\rm p}=300$ is required to measure halo shapes and orientations accurately, i.e. to have less than $5\,\%$ deviation of the measured axis ratios from the truth, and less than $5\,$degrees uncertainty in the direction of the largest eigenvector \citep[see also][]{bett07}. Sparsely sampled haloes tend to produce smaller axis ratios (and hence larger ellipticity on average) as well as rapidly increasing uncertainty in the halo orientation, leading to an underestimation of ellipticity correlations \citep{jing02}.

The minimum requirement can be relaxed to $N_{\rm p}=100$ when accepting $10\,\%$ deviation of the measured axis ratios and $10\,$degrees deviation of the measured orientation of the largest eigenvector. Since $N_{\rm p} \geq 300$ is a restrictive condition that would discard more than half of the haloes identified in the Millennium Simulation, we will adopt this less stringent limit when comparing the simulation intrinsic alignment signals to data and when predicting II and GI signals, but keep to a minimum particle number of 300 for the investigation of dependencies in the simulation of intrinsic alignment correlations in Section$~$\ref{sec:simresults}.

Furthermore we make use of the \citet{bett07} calculations of the specific angular momentum of haloes within the virial radius, restricted to $N_{\rm p} \geq 300$. We find that shape models based on the halo angular momentum have negligible correlations for halo masses slightly above $300 m_{\rm p}$ (see Section$~$\ref{sec:simresults}), so that the absence of angular momentum information for low-mass haloes should not affect our results. 

We test whether our results depend on the radius within which the angular momentum is calculated. To this end, we calculate correlations between the normalised angular momenta of massive haloes with at least 300 particles within $0.1 r_{\rm vir}$. However, the correlations in orientation resulting from computing $\vek{L}$ within $0.1 r_{\rm vir}$ and within $r_{\rm vir}$ do not show significant deviations beyond the error bars computed from field-to-field variance.

\subsection{Early-type galaxies}
\label{sec:Emodels}

All central galaxies with $R_{\rm type} \geq 0.6$ and in haloes with more than 100 particles are assumed to have the same three-dimensional shape as their host haloes. The complex galaxy ellipticity (in projection) is computed from the symmetric tensor ${\mathbf W}$ as defined in Section 3.1 of Paper I via\footnote{Note that in this paper we consistently work in terms of the complex ellipticity $\epsilon$, as opposed to the polarisation $e$ which was employed in Paper I. The two quantities are related via $e = 2 \epsilon/(1+ |\epsilon|^2)$ \citep{bartelmann01}.}
\eqa{
\label{eq:ellipseprojection3}
\epsilon_1 &=& \frac{W_{11}-W_{22}}{W_{11}+W_{22}+2 \sqrt{\det {\mathbf W}}}\;;\\ \nn
\epsilon_2 &=& \frac{2\, W_{12}}{W_{11}+W_{22}+2 \sqrt{\det {\mathbf W}}}\;.
}

A frequently used alternative is to base the shape of elliptical galaxies on the reduced inertia tensor of the halo, arguably producing a better approximation of the shape of the galaxy residing close to the halo centre as it gives more weight to small radii. To model the impact of switching from the simple inertia tensor of Equation (\ref{eq:massquadrupole}) to the reduced one, we devise models for which we increase the halo semi-axis ratios by about $25\,\%$, thereby approximating the results of \citet{bett11}. Early-type models are denoted by \texttt{Est} if based on the simple inertia tensor, and by \texttt{Ert} if based on the reduced inertia tensor.

Additionally we construct a model named \texttt{Ema} which is based on the simple inertia tensor and includes a random misalignment of halo major axes. The misalignment angles are drawn from a Gaussian distribution with a scatter of $35^\circ$, as proposed by \citet{okumura08} who obtained this scatter by matching the intrinsic ellipticity correlations of an SDSS LRG sample to halo shape correlations extracted from N-body simulations.

Since ellipticity correlations among early-type galaxies, whose haloes have $N_{\rm p} < 100$ and thus no measured halo shape, are expected to be small (see also Section$~$\ref{sec:simresults}), it is safe to assume that these galaxies have random orientations. We make the further assumption that the statistical halo shape properties of galaxies with $N_{\rm p} < 100$ are the same as those of more massive galaxies. In each redshift slice we construct two-dimensional histograms of halo axis ratios and assign randomly sampled shapes from these histograms to low-mass galaxies at the same redshift.

\subsection{Late-type galaxies}
\label{sec:Smodels}

All central galaxies with $N_{\rm p} \geq 300$ and $R_{\rm type} < 0.6$ are modelled as circular thick discs whose orientation is determined by the angular momentum vector $\vek{L}=\bc{L_x,L_y,L_\parallel}^\tau$ of the underlying halo. The ellipticity of the galaxy image is determined analogously to the procedure outlined in Paper I, except that, due to the different definition of ellipticity, Paper I, Equation (6) is modified to
\eqa{
\label{eq:spiraleps1}
\epsilon_1 &=& \cos (2 \theta)\; \frac{1-r}{1+r}\;;\\ \nn
\epsilon_2 &=& \sin (2 \theta)\; \frac{1-r}{1+r}\;,
}
where $\theta$ is the polar angle of the image ellipse. The axis ratio of the ellipse reads
\eq{
\label{eq:spiraleps3}
r = \frac{|L_\parallel|}{|\vek{L}|} + r_{\rm edge-on}\; \sqrt{1 - \frac{L_\parallel^2}{|\vek{L}|^2}  }\;,
}
where $r_{\rm edge-on}$ is the ratio of disc thickness to disc diameter, i.e. approximately the axis ratio for a galaxy viewed edge-on. We explore two values of this parameter: $r_{\rm edge-on}=0.1$, similar to what is expected for the disc of e.g. the Milky Way (models denoted by \texttt{Sth}), and $r_{\rm edge-on}=0.25$, designed to include the effect of bulges (models denoted by \texttt{Sal}) and motivated by the observations of \citet{bailin08}.

Like in the case of the early-type galaxy shape model, the projection implicitly assumes that the three-dimensional light distribution is uniform with a sharp cut-off at the perimeter. We refrain from using more complicated schemes involving a realistic radial light distribution as this could imply variable ellipticity as a function of radius. Moreover we neglect any small deviations of the image from an elliptical shape in the projection of discs.

Re-simulations of dark matter haloes with high resolution and including baryonic particles modelled in different ways indicate that galaxy discs are not perfectly aligned with the halo angular momentum. \citet{bett11} collected the misalignment results from over 500 haloes simulated with baryons and galaxy formation physics (from \citealp{deason11} and \citealp{bett10}, using the simulations of \citealp{crain09} and \citealp{okamoto05} respectively), as well as 95 haloes without baryons (also from \citealp{bett10}). \citet{bett11} found that the polar misalignment angles $\vartheta$ between angular momentum vector and disc rotation axis can be jointly described by the distribution
\eq{
\label{eq:Lmisalignment1}
P(\cos \vartheta) = \frac{1}{2 \sigma^2 \sinh \sigma^{-2}}\; \exp \br{ \frac{\cos \vartheta}{\sigma^2} }\;,
}
with $\sigma=0.55$. The cumulative distribution function of equation (\ref{eq:Lmisalignment1}) is conveniently calculated and inverted analytically, so that we can compute misalignment angles via
\eq{
\label{eq:Lmisalignment2}
\vartheta = \arccos \bc{ \sigma^2 \ln \bb{ \expo{-\sigma^{-2}} + 2 \sinh \sigma^{-2} u }}\;,
}
where $u$ is a uniform random number in $\bb{0;1}$. We determine a new vector $\widetilde{\vek{L}}$ by rotating $\vek{L}$ by an amount $\vartheta$ around an arbitrary orthogonal axis, followed by another rotation around $\vek{L}$ by an angle $\beta$ uniformly sampled in $\bb{0;2\pi}$. The galaxy disc is then constructed perpendicular to $\widetilde{\vek{L}}$. 

Note that \citet{heymans04,heymans06} used a similar prescription for late-type galaxy models. Disc thickness is accounted for by rescaling ellipticities as $\epsilon_{\rm gal} = 0.73\, \epsilon_{\rm thin~disc}$, which in the edge-on limit corresponds to an axis ratio of 0.16, hence lying in-between the models for $r_{\rm edge-on}$ that we consider. Misalignment angles between angular momentum vector and rotation axis were extracted from the simulations by \citet{bosch02}. The set of simulations studied in \citet{bett11} is superior in number of galaxy/halo systems and resolution, although both agree on a median misalignment of the order $30\,$degrees. The fit to the distribution of $\vartheta$ suggested by \citet{bett11} appears to be more accurate in particular for large misalignments than the truncated Gaussian proposed by \citet{heymans04}. This misalignment prescription is implemented for disc models with $r_{\rm edge-on}=0.25$, denoted as \texttt{Sma}.

Analogous to low-mass early-type galaxies, late-type galaxies with $N_{\rm p} < 300$ that have no angular momentum information are modelled as randomly oriented thick discs, where $r_{\rm edge-on}$ has the same value as the model used for the corresponding model of central late-type galaxies with $N_{\rm p} \geq 300$.

\subsection{Satellite galaxies}
\label{sec:satmodels}

For galaxies residing in the substructures of haloes we do not have information about the properties of their dark matter distribution. Therefore we have to rely on external data to model both the shapes and orientations of satellite galaxies. 

For early-type satellites we proceed in analogy to the central galaxies with $N_p<100$ and sample the axis ratios of three-dimensional ellipsoids from the histograms obtained for massive haloes with shape information at each redshift slice (model \texttt{est}). Optionally these axis ratios are rescaled to mimic the use of the reduced inertia tensor (model \texttt{ert}). The ellipsoids are then oriented to point their major axis towards the central galaxy of the halo and subsequently projected along the line of sight using Equation (\ref{eq:ellipseprojection3}) to yield image ellipticities.

The radial alignment of satellite galaxies towards the centre of the halo is strongly supported by simulations \citep[e.g.][]{pereira08,knebe08,knebe10} and has also been detected in galaxy clusters \citep[e.g.][]{pereira05}, although the statistical significance is still low and detections are plagued by systematic effects \citep[see][]{hao11}. Simulation results indicate that tidal torquing is the cause for satellite alignment \citep{pereira08,knebe10}. The perfect radial alignment that we assume by default is to be understood as an upper limit for the satellite ellipticity correlation signal.

As an alternative model we implement the modifications of shapes and orientations of subhaloes found by \citet{knebe08} in high-resolution dark matter-only simulations (Knebe08 model hereafter; identifier \texttt{ekn}). The corresponding modifications to halo shapes are detailed in Paper I, Section 3.3. The same publication also confirmed that satellite galaxies are radially aligned towards their host, albeit with substantial scatter in the angle $\vartheta'$ between the satellite major axis and the direction to the halo centre. The probability distribution of $\vartheta' \in \bb{0;\pi/2}$ is given by (\citealp{knebe08}; typo corrected)
\eq{
\label{eq:knebedistribution}
P(\cos \vartheta') = \frac{A \cos^4 \vartheta' + B}{A/5 + B}\;,
}
where the constants $A \approx 2.64$ and $B \approx 0.59$ are independent of the halo mass. After changing the satellite major axis orientation by an angle $\vartheta'$ sampled from Equation (\ref{eq:knebedistribution}), we allow for a rotation by $\beta$, uniformly sampled in $\bb{0;2\pi}$, around the connecting line to the halo centre, before projecting the satellite shape along the line of sight.

Similar to their central counterparts, late-type satellite galaxies are assumed to be thick circular discs perpendicular to the angular momentum vector of the halo, where we place the angular momentum vector at an arbitrary pointing perpendicular to the line connecting the position of the satellite with the centre of the halo. This means that a satellite disc galaxy viewed edge-on \lq points\rq\ towards the centre of a halo. Otherwise we explore the same model variants as for central late-type galaxies, i.e. the two values of $r_{\rm edge-on}$ (models \texttt{sth} and \texttt{sal} for $r_{\rm edge-on}=0.1$ and $r_{\rm edge-on}=0.25$ respectively) and the optional misalignment of the angular momentum vector from its default direction according to \citet{bett11}, which in this case is equivalent to the misalignment of the position vector of the satellite with the major axis of the projected image of an inclined disc (model \texttt{sma}).

\section{Simulated galaxy intrinsic shape correlations}
\label{sec:simresults}

We begin our analysis by studying galaxy shape correlations and their trends with various galaxy properties, separated into early types (i.e. based on halo shapes) and late types (i.e. based on halo angular momenta). Here we focus on correlations of the projected shapes of two galaxies as a function of their three-dimensional, comoving separation $r$, computed via \citep[e.g.][]{heymans06}
\eq{
\label{eq:3Dcorr}
\eta(r) = \ba{ \epsilon_+(\vek{x})\, \epsilon_+(\vek{x}+\vek{r}) }_{\svek{x}} +  \ba{ \epsilon_\times(\vek{x})\, \epsilon_\times(\vek{x}+\vek{r}) }_{\svek{x}}\;,
}
where $\ba{\cdot}_{\svek{x}}$ denotes the average over all positions $\vek{x}$. The tangential(+) and cross($\times$) components of the galaxy ellipticity are defined as \citep{bartelmann01}
\eqa{
\label{eq:tangentialcross}
\epsilon_+ &=& -\epsilon_1 \cos(2 \varphi) - \epsilon_2 \sin(2 \varphi)\;;\\ \nn
\epsilon_\times &=& \epsilon_1 \sin(2 \varphi) - \epsilon_2 \cos(2 \varphi)\;,
}
where $\varphi$ denotes the polar angle of the line connecting the galaxy pair in a reference coordinate system. The statistic $\eta(r)$ allows for a compromise between the proximity to observable quantities (which are projected along the line of sight) and the readiness with which the signal can be physically interpreted. 

We perform the measurement on the simulation boxes, choosing the line of sight parallel to the edges of the cube. Error bars are determined from the variance between eight equal-sized cubic sub-volumes of the simulation box. Correlations between measurements of $\eta(r)$ at different separation $r$ are generally weak, the correlation coefficient exceeding 0.5 only occassionally.

Unless explicitly stated otherwise, we will limit ourselves in this section to galaxies for which we have reliable measurements of halo properties, i.e. central galaxies with halo particle number exceeding 300. Early-type models are based on the simple inertia tensor, and for late-type models $r_{\rm edge-on}=0.25$ and zero misalignment between disc rotation axis and angular momentum vector is assumed. In Section$~$\ref{sec:modeleffect} we will highlight the impact of different assumptions for our central galaxy models and the effect of including satellite galaxies.

\subsection{Redshift dependence}
\label{sec:zdep}

\begin{figure}
\centering
\includegraphics[scale=.34,angle=270]{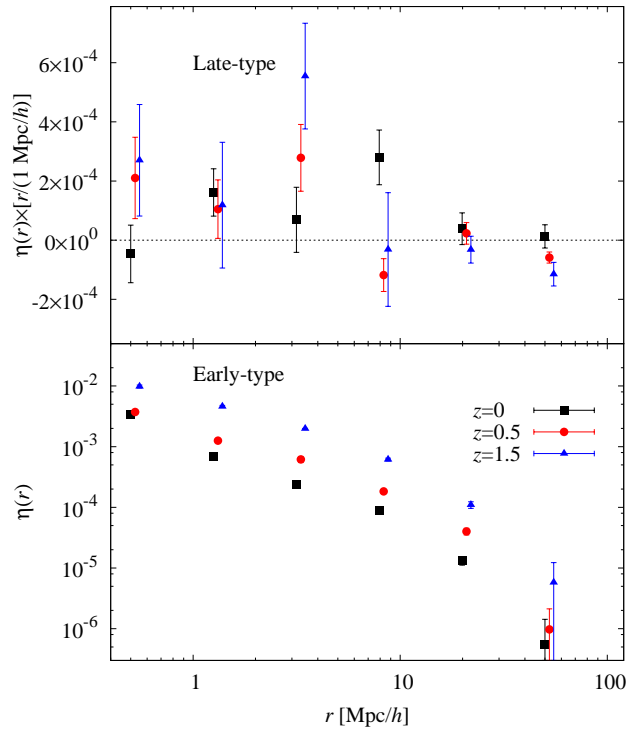}
\caption{Correlation function $\eta(r)$ as a function of comoving galaxy pair separation $r$ for late-type (top panel) and early-type (bottom panel) galaxies. Black squares show the results for galaxies at redshift $z=0$, red circles for galaxies at $z=0.5$, and blue triangles for galaxies at $z=1.5$. The sample is restricted to the mass range $3.5 \times 10^{11}\,M_\odot<m_{\rm halo}<3.5 \times 10^{12}\,M_\odot$. Note the different scaling of the ordinate axes.}
\label{fig:3D_z}
\end{figure} 

\begin{figure}
\centering
\includegraphics[scale=.34,angle=270]{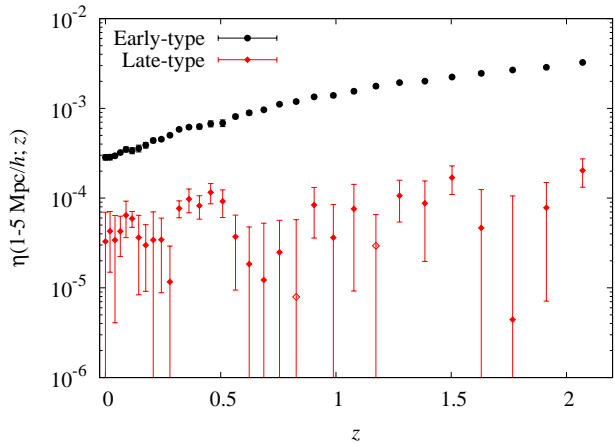}
\caption{Correlation function $\eta(r)$ computed for every redshift output in a single radial bin covering $1\,{\rm Mpc}/h<r<5\,{\rm Mpc}/h$. Results for early-type (late-type) galaxies are shown as black dots (red diamonds). Negative points are plotted with their absolute values as open symbols. Since the signal is computed from the same simulation volume at different redshifts, points are strongly correlated.}
\label{fig:3D_z_analysis}
\end{figure}
 
The dependence of the galaxy ellipticity correlation function on redshift is illustrated in Fig.$\,$\ref{fig:3D_z}, where $\eta(r)$ is plotted for three redshift outputs, at $z=0$, $z=0.5$, and $z=1.5$, respectively. The samples are restricted to the halo mass range $3.5 \times 10^{11} M_\odot<m_{\rm halo}<3.5 \times 10^{12} M_\odot$ to avoid confusion with a potential mass dependence of the correlations. 

Late-type galaxies generally do not show significant correlations of their projected ellipticities. The two high-redshift samples have marginally positive correlations below $r \sim 5\,{\rm Mpc}/h$, while at $z=0$ $\eta(r)$ is consistent with a constant small positive signal on all scales. To achieve higher signal-to-noise, we compute $\eta(r)$ in a single radial bin covering $1\,{\rm Mpc}/h<r<5\,{\rm Mpc}/h$, plotted as a function of redshift in Fig.$\,$\ref{fig:3D_z_analysis}. For late-type galaxies the resulting correlations are marginally positive at most output redshifts, increasing slightly in amplitude by a factor of a few between $z=0$ and $z=2$.

Ellipticity correlations for early-type galaxies are statistically significant out to $r \ga 20\,{\rm Mpc}/h$ for all redshift bins, increasing by about three orders of magnitude from $r = 50\,{\rm Mpc}/h$ to $r = 0.5\,{\rm Mpc}/h$ (see Fig.$\,$\ref{fig:3D_z}). The overall amplitude of $\eta(r)$ decreases towards lower redshifts although this decrease is weaker on non-linear scales, i.e. for $r \la 1\,{\rm Mpc}/h$ at $z=0.5$ and $r \la 2\,{\rm Mpc}/h$ at $z=0$. This may be caused by an increase of alignments in regimes of strongly non-linear structure growth. The decline of correlation strength is also evident from Fig.$\,$\ref{fig:3D_z_analysis}, and is, to good approximation, exponential as a function of look-back time (and hence spacing of output redshifts). The only exception to this trend is a mild excess signal at $z <0.1$ which we relate to the larger impact of non-linear scales in the range $1\,{\rm Mpc}/h<r<5\,{\rm Mpc}/h$ at these redshifts.

While the mass cuts remove any redshift dependence of the mean halo mass for the late-type sample, there is still a moderate increase by $28\,\%$ from $z=1.5$ to $z=0$ for the early-type sample. Since correlations are expected to become stronger with increasing halo mass (see the following section), this might lead to a small under-estimation of the redshift dependence. Note that, since we fix the halo mass, we are not tracking the same sample of haloes as a function of redshift. For instance, the high-redshift haloes with strong alignments correspond to objects that formed early and are likely to be among the most massive objects at $z=0$. The results of Fig.$\,$\ref{fig:3D_z_analysis} thus suggest a link between an early formation time of the halo and strong intrinsic alignments.

To gain more insight into the evolution of galaxy ellipticity correlations with redshift, it is useful to consider the alignment between the major axes (normalised to unit-length, denoted by $\vek{s}_1$) for the same galaxy samples, as this measure employs a three-dimensional quantity and thus avoids projection effects. We compute the correlation function
\eq{
\eta_{\rm 3D}(r) = \ba{\vek{s}_1(\vek{x}) \cdot \vek{s}_1(\vek{x}+\vek{r}) }_{\svek{x}}\;-0.5,
}
where 0.5 is subtracted to ensure that random orientations of the major axes imply a null signal.

The statistic $\eta_{\rm 3D}(r)$ features the same scaling with redshift as shown for $\eta(r)$ in the bottom panel of Fig.$\,$\ref{fig:3D_z} (hence not shown). Similarly, \citet{lee08} observed an increase in the correlation of halo major axes at high redshift in the Millennium Simulation, although they used different halo definitions and selection criteria. Strong intrinsic alignment signals at high redshift pose a challenge to deep weak lensing surveys, in particular if a similar increase happens for correlations between galaxy ellipticity and the mass distribution (and hence the gravitational shear). That such correlations are indeed significant is suggested by the results of \citet{lee08} who found an increase of halo major axis-halo number density correlations with redshift.

\subsection{Mass and luminosity dependence}
\label{sec:massdep}

\begin{figure}
\centering
\includegraphics[scale=.34,angle=270]{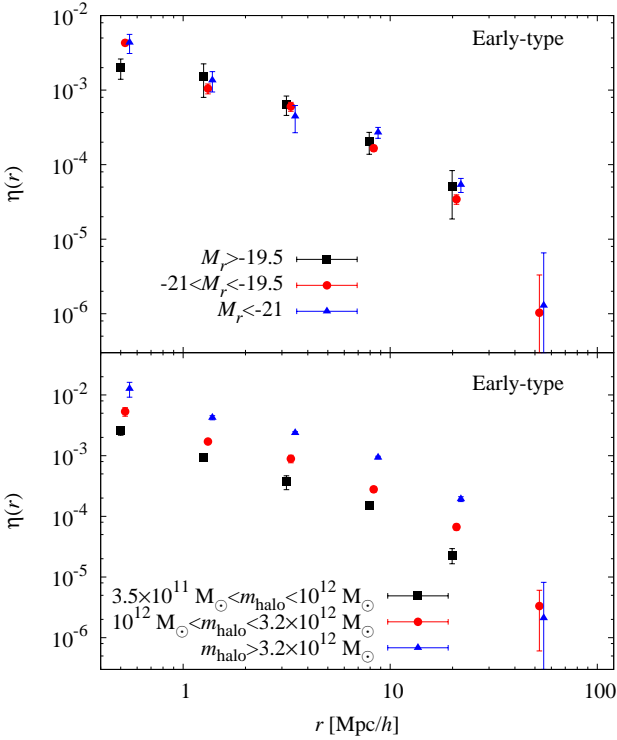}
\caption{\textit{Top panel}: Correlation function $\eta(r)$ as a function of comoving galaxy pair separation $r$, split into three magnitude bins (rest frame, $r$-band). Shown are correlations between early-type galaxies at $z=0.5$ in the halo mass range $3.5 \times 10^{11} M_\odot < m_{\rm halo} < 3.5 \times 10^{12} M_\odot$ for $M_r>-19.5$ ($-21 < M_r < -19.5$; $M_r < -21$) as black squares (red dots; blue triangles). \textit{Bottom panel}: Same as above, but split into three halo mass bins. Correlations between early-type galaxies at $z=0.5$  in the mass range $3.5 \times 10^{11} M_\odot < m_{\rm halo} < 10^{12} M_\odot$ ($10^{12} M_\odot < m_{\rm halo} < 3.2 \times 10^{12} M_\odot$; $m_{\rm halo} > 3.2 \times 10^{12} M_\odot$) are shown as black squares (red dots; blue triangles).}
\label{fig:3D_mass}
\end{figure} 

\begin{figure}
\centering
\includegraphics[scale=.33,angle=270]{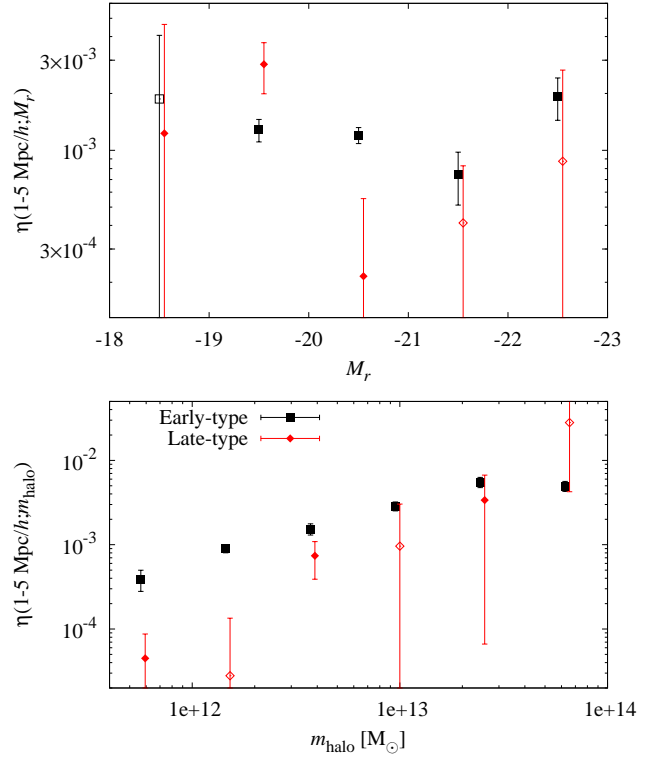}
\caption{Correlation function $\eta(r)$ at $z=0.5$ in a single radial bin covering $1\,{\rm Mpc}/h<r<5\,{\rm Mpc}/h$ as a function of halo mass $m_{\rm halo}$ (bottom panel) and rest-frame magnitude $M_r$ at fixed mass (top panel). The signal for early-type (late-type) galaxies is shown as black squares (red diamonds).  Negative points are plotted with their absolute values as open symbols.}
\label{fig:3D_mass_analysis}
\end{figure} 

An increase of galaxy number density-ellipticity correlations with luminosity for early types was firmly established in SDSS LRG samples by \citet{hirata07} and confirmed with a substantially extended data set by \citet{joachimi11}. It is reasonable to assume that this result implies stronger intrinsic correlations and stronger alignments with the large-scale structure of more massive dark matter haloes. 

We test this scenario with our mock sample by calculating $\eta(r)$ at $z=0.5$ for three mass and rest-frame magnitude bins, respectively, as shown in Fig.$\,$\ref{fig:3D_mass}. To cleanly isolate the mass and luminosity dependencies, we restrict the analysis of the luminosity dependence to galaxies in the mass range $3.5 \times 10^{11}\,M_\odot<m_{\rm halo}<3.5 \times 10^{12}\,M_\odot$. Note that the results for late-type galaxies are noisy and without clear signal detections, so that we only show early-type correlations in this plot.

We observe a clear increase in the overall amplitude of $\eta(r)$ for the higher mass samples (bottom panel of Fig.$\,$\ref{fig:3D_mass}), without significant changes in the radial dependence. The top panel of Fig.$\,$\ref{fig:3D_mass} shows no clear dependence of intrinsic correlations on luminosity for fixed mass range. We re-compute $\eta(r)$ with narrower bins in $m_{\rm halo}$ and $M_r$, using only a single radial bin in the range $1\,{\rm Mpc}/h<r<5\,{\rm Mpc}/h$. Moreover, to avoid any residual effect of the pronounced mass dependence, we choose the mass ranges for each $M_r$ bin to cover a decade in $m_{\rm halo}$ and have an average of $m_{\rm halo} \approx 2 \times 10^{12}\,M_\odot$.

Figure \ref{fig:3D_mass_analysis}, top panel, reveals that there is no clear luminosity dependence of $\eta(r)$ if the halo mass is kept fixed. The amplitude of the early-type correlations as a function of $m_{\rm halo}$ follows a power-law up to $m_{\rm halo} \approx 4 \times 10^{13} M_\odot$. Only the point at the highest masses, which includes the haloes of groups and clusters, drops below this relation. This could possibly indicate that the high merger activity expected for these objects reduces the halo ellipticity on average and/or partially destroys halo alignments. \citet{lee08} reported a similarly strong mass dependence of major axis correlations using two mass bins over a similar mass range as in Fig.$\,$\ref{fig:3D_mass_analysis}. Late-type galaxies generally show a very weak signal, with no clear trends seen in the mass and luminosity dependence.

The results of Fig.$\,$\ref{fig:3D_mass_analysis} demonstrate that mass rather than luminosity is the dominant parameter governing the intrinsic alignment strength. In flux-limited samples, on which weak lensing studies are usually based, the combined redshift and mass dependencies lead to an even more pronounced increase of the alignment signal with redshift at a given physical galaxy separation. However, as a function of angular scale, this scaling is counteracted by the drop in intrinsic alignment strength for pairs of galaxies with larger physical separation.

\subsection{Environment dependence}
\label{sec:densitydep}

\begin{figure}
\centering
\includegraphics[scale=.34,angle=270]{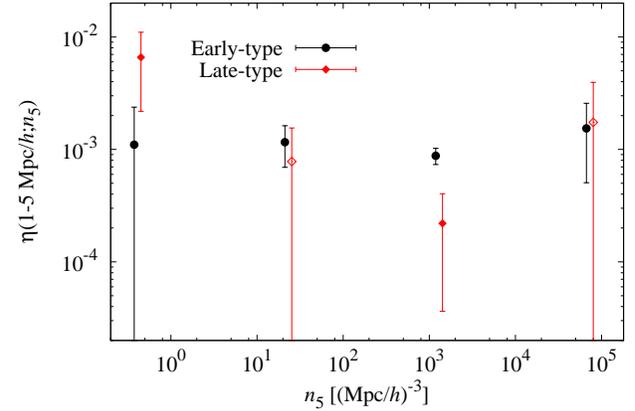}
\caption{Correlation function $\eta(r)$ at $z=0.5$ and for fixed halo mass in a single radial bin covering $1\,{\rm Mpc}/h<r<5\,{\rm Mpc}/h$ as a function of local galaxy number density $n_5$. The signal for early-type (late-type) galaxies is shown as black dots (red diamonds). Negative points are plotted with their absolute values as open symbols.}
\label{fig:3D_ndensdep}
\end{figure} 

An estimate of the local density around a given galaxy is determined by computing the comoving distance $d_5$ to the fourth-nearest neighbour, where satellite galaxies are included, and deriving $n_5 \equiv 5/[(4/3) \pi d_5^3]$. We refrain from calculating mass densities because we only have halo masses for central galaxies, and besides, number densities are more readily constrained observationally. 

As we do not find significant differences in amplitude or radial dependence between subsamples binned according to $n_5$, we again compute $\eta(r)$ for a single radial bin in the range $1\,{\rm Mpc}/h<r<5\,{\rm Mpc}/h$, devising four logarithmically spaced bins in number density. Like in the preceding section, we minimise the impact of a possible residual mass dependence by setting the mass range of haloes included in each $n_5$ bin to span a decade and have a mean of $m_{\rm halo} \approx 2 \times 10^{12}\,M_\odot$.

As shown in Fig.$\,$\ref{fig:3D_ndensdep}, the amplitude of early-type correlations is constant across more than five orders of magnitude in $n_5$. The signal for late-type galaxies is again noisy, with a marginal preference for stronger alignments in the lowest density bin. We deem it physically plausible that it is easier for isolated spiral galaxies to remain aligned after formation with the surrounding large-scale structure, and thus neighbouring galaxies in a similar environment, than for disc-dominated galaxies in high-density regions which frequently and/or strongly interact. The spheroidal galaxies in our early-type sample have all undergone at least one major merger, so that in this case a different, apparently density-independent alignment mechanism, which is preserved or possibly generated by strong gravitational interactions, must be at work.

A dependence of galaxy shape correlations on environment could complicate attempts to model intrinsic alignments and cause subtle biases when trying to infer the properties of the large-scale structure via weak lensing observations, making an observational test of these results highly desirable. The only attempt hitherto to constrain environment effects on intrinsic alignment signals was undertaken by \citet{hirata07} who isolated the brightest galaxies of groups and clusters in the SDSS LRG sample. However, this selection more closely resembles a split into central and satellite galaxy populations and is therefore not directly compatible to our results.

\subsection{Effect of modelling assumptions}
\label{sec:modeleffect}

\begin{figure}
\centering
\includegraphics[scale=.34,angle=270]{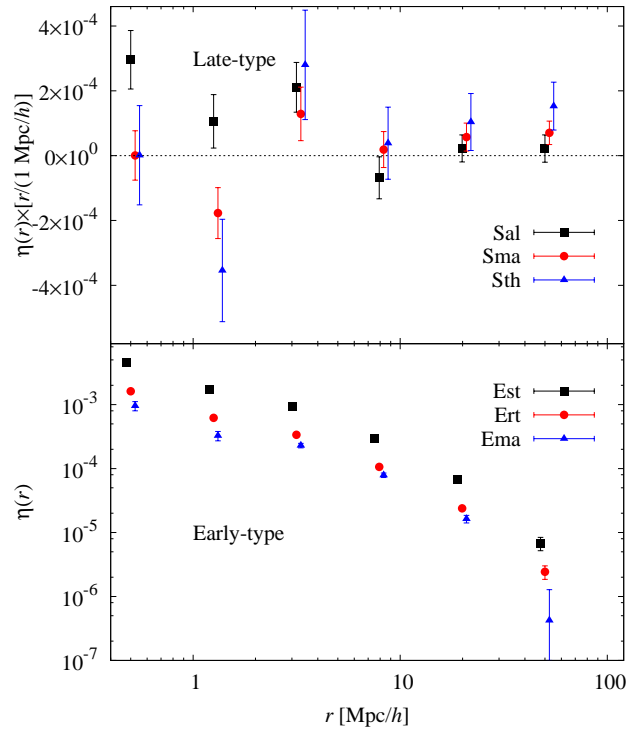}
\caption{Correlation function $\eta(r)$ as a function of comoving galaxy pair separation $r$ for different modelling assumptions about shapes of central galaxies, in the redshift range $0.4<z<0.6$. \textit{Top panel}: Results for late-type galaxies modelled as perfectly aligned with the angular momentum of their host halo (black squares), modelled as having a random misalignment between disc rotation axis and halo angular momentum vector (red circles), and modelled as in the foregoing case but with a thinner disc of $r_{\rm edge-on}=0.1$ instead of $r_{\rm edge-on}=0.25$ (blue triangles). \textit{Bottom panel}: Results for early-type galaxies with shapes based on measurements of the simple inertia tensor (black squares), the reduced inertia tensor (red circles), or including random misalignments (blue triangles). Note the different scaling of the ordinate axes.}
\label{fig:3D_models}
\end{figure} 

So far we have only used central haloes with $N_P \geq 300$ for which the galaxy models are derived directly from the halo properties measured in the simulation. Furthermore we have assumed perfect alignment between disc rotation axes and angular momentum vectors, as well as halo shapes and galaxy shapes, to obtain the clearest correlation signals. In Fig.$\,$\ref{fig:3D_models} we show how the different assumptions made about the galaxy models change the results presented in Section$~$\ref{sec:galaxymodels}. Note that in this section we measure the correlations in the light cones (which include satellite galaxies) rather than the simulation boxes, limiting the redshift range to $0.4<z<0.6$, and considering haloes with $N_P \geq 300$ only. Error bars are now and henceforth determined from the field-to-field variance in the 64 light cones.

If galaxy shapes are assigned according to the reduced inertia tensor measurements (model \texttt{Ert}), $\eta(r)$ is suppressed by a factor of 2.8, equally across all scales, as all projected galaxy images are closer to circular. The interiors of haloes, where the reduced inertia tensor has higher weight, are not only closer to spherical but are also less aligned with the large-scale structure \citep[e.g.][]{schneiderm11} and hence with each other across large distances. This radial dependence of halo alignment would further reduce the amplitude of correlations, manifesting the strong dependence of shape correlations on the details of the method with which simulated dark matter halo shapes are measured. Note that, nonetheless, correlations for early-type galaxies are significant out to around $40\,{\rm Mpc}/h$.

Random misalignments (model \texttt{Ema}) lead to a suppression of the amplitude of $\eta(r)$ by about a factor of 4.5, in good agreement with \citet{okumura08}. The rescaling is slightly dependent on physical separation, with stronger suppression below $1\,{\rm Mpc}/h$.

\begin{figure}
\centering
\includegraphics[scale=.34,angle=270]{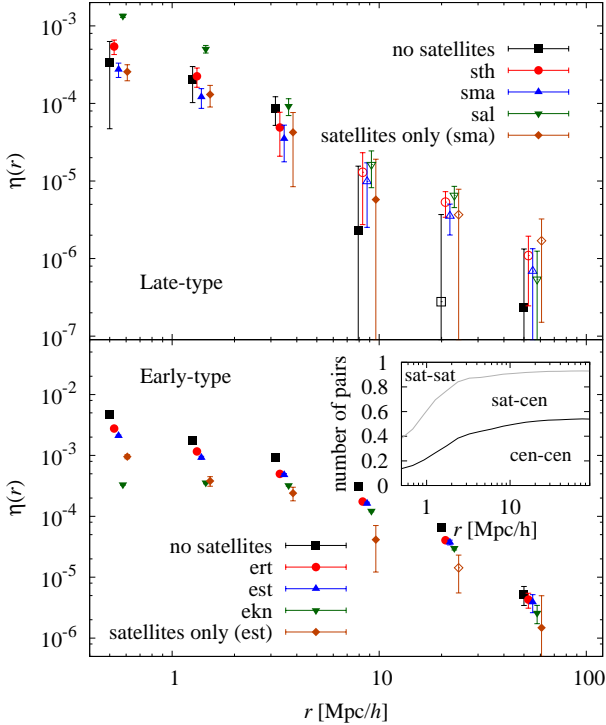}
\caption{Correlation function $\eta(r)$ as a function of comoving galaxy pair separation $r$ for different contributions of satellite galaxies. Galaxies are selected to lie in the range $0.4<z<0.6$ and be brighter than $M_r<-20$. \textit{Top panel}: Results for late-type galaxies, where black squares correspond to the signal without satellite galaxies, red circles (blue upward triangles) to modelling satellites as misaligned discs with $r_{\rm edge-on}=0.1$ ($r_{\rm edge-on}=0.25$), green downward triangles to satellites which are perfectly aligned with the direction towards the central galaxy, and brown diamonds to the signal for staellites only. Negative points are shown as open symbols. \textit{Bottom panel}: Results for early-type galaxies, where black squares correspond to the signal without satellite galaxies, red circles (blue upward triangles) to modelling satellite shapes based on the reduced (simple) inertia tensor, green downward triangles to satellites modelled according to the Knebe08 model, and brown diamonds to satellites only. The inset shows the fraction of central-central (cen-cen), satellite-central (sat-cen), and satellite-satellite (sat-sat) galaxy pairs contributing to the correlations at different $r$. Note that differences in pair composition between early- and late-type samples are small.}
\label{fig:3D_sat}
\end{figure} 

All our late-type models are consistent with zero over the range shown in Fig.$\,$\ref{fig:3D_models}, where the models \texttt{Sma} and \texttt{Sth} have a $p$-value that is about twice as high as for model \texttt{Sal}. It is expected that random misalignments of the angular momentum vector (model \texttt{Sma}) further dilute any signal. Using a thinner disc while retaining the misalignments (model \texttt{Sth}) moves data points marginally away from zero but also increases the scatter and hence the error bars.

Returning to the default models for central galaxies but now including satellites in $\eta(r)$, we obtain the correlation functions shown in Fig.$\,$\ref{fig:3D_sat}. Satellite galaxies modify the correlation signal most significantly on scales below about $3\,{\rm Mpc}/h$ where they clearly dominate the galaxy pairs available (see the inset). The models \texttt{sth}, \texttt{sma}, \texttt{ert}, and \texttt{est} have similar correlation signals as the correponding central galaxies (cf. the respective plots for satellite galaxies only in Fig.$\,$\ref{fig:3D_sat}) and thus cause little change in the total $\eta(r)$.  As expected, satellite models with strong alignment like \texttt{sal} boost the signal on small scales while models with less elliptical shapes and weaker alignments (\texttt{ekn}) yield the strongest suppression of $\eta(r)$.

\section{Comparison with intrinsic alignment observations}
\label{sec:iaobs}

\subsection{Early-type samples \& method}
\label{sec:megazmethod}

\begin{figure*}
\centering
\includegraphics[scale=.36,angle=270]{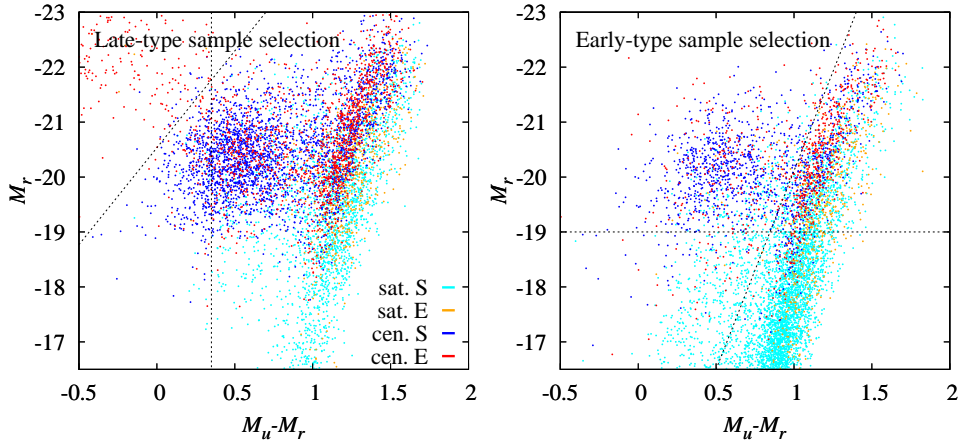}
\caption{Rest-frame colour-magnitude diagrams of the simulation-based galaxy samples. \textit{Left panel}: Galaxies are selected to match the WiggleZ sample, by applying the cut $20 < r < 21.8$. Early-type central (satellite) galaxies are assigned red (orange) points; late-type central (satellite) galaxies blue (cyan) points. The black dotted lines indicate the cuts used to isolate the blue part of the blue cloud and suppress the contamination by bright blue galaxies classified as early types. Note that we treat S0 galaxies as late-types in this work. \textit{Right panel}: Same as on the left, but with galaxies selected to match the early-type SDSS samples, i.e. using the cuts $z<0.7$ and $M_r<-19$. The black dotted lines indicate the cuts used to isolate the bright red sequence; only galaxies above the horizontal line and to the right of the vertical line are included. For easier inspection we only plot every 10th (50th) galaxy for the WiggleZ (early-type) sample.}
\label{fig:cmd}
\end{figure*} 

We now turn to the comparison of two-point statistics of galaxy shapes from the Millennium Simulation with the results from observational data sets. To test our models of early-type galaxy shapes, we will use the most comprehensive analysis to date by \citet{joachimi11}, based on several early-type SDSS samples. Relying on the shape measurements by \citet{mandelbaum06}, \citet{joachimi11} jointly analysed early-type galaxy samples constructed from the SDSS Main and Luminous Red Galaxy (LRG) spectroscopic samples, and the MegaZ-LRG sample \citep{collister07} which features photometric redshifts based on SDSS $ugriz$ photometry. 

The SDSS Main and MegaZ-LRG samples were each split into two redshift bins chracterised by a substantially different mean luminosity. The LRG sample was divided into three luminosity bins plus a further split in redshift, so that in total 10 data sets were analysed, with quite uniform coverage of redshifts out to $z \la 0.7$ and rest-frame magnitudes\footnote{Note that rest-frame magnitudes of the observational data sets were determined via the $k+e$-corrections of \citet{wake06}, but using the same values of the relevant cosmological parameters as those of the Millennium Simulation.} over the range $-23 < M_r < -19$. High-quality shape measurements are available for about $74,000$ galaxies with spectroscopic redshifts and more than $400,000$ galaxies with photometric redshifts.

The selection criteria of the different samples are summarised in \citet{joachimi11}. If we applied these directly to the simulated galaxy catalogues, the analysis would be subjected to the potentially unreliable colours produced by semi-analytic galaxy evolution models; see e.g. \citet{cohn07} who found that the red sequence extracted from the Millennium Simulation has a tilt and larger spread compared to observations (note that they did not use a \texttt{GALFORM}-based model though). Hence we refrain from applying survey selection criteria that involve colours directly to the simulation-based catalogues. Moreover the luminous galaxies in the SDSS samples have low number densities and cover about one order of magnitude larger survey area than the combined Millennium catalogues, so that the selected mock samples would be undesirably small.

Hence we chose a different route: after homogenising the colours of the samples, \citet{joachimi11} jointly fit a three-parameter model, finding excellent agreement between every data set and the model. It is based on the linear alignment model \citep{hirata04,hirata10}, heuristically extended into the non-linear regime \citep{bridle07}, and fitted with extra redshift and luminosity dependencies, which results in the matter-intrinsic power spectrum (for a formal definition see the appendix of \citealp{joachimi11})
\eqa{
\label{eq:IAmodel}
P^{\rm model}_{\delta {\rm I}}(k,z,L) &=& -A\; C_1\; \rho_{\rm cr} \frac{\Omega_{\rm m}}{D(z)}\; P_{\delta}(k,z)\\ \nn
&& \hspace*{0.3cm} \times\; \br{\frac{1+z}{1+z_0}}^{\eta_{\rm other}} \br{\frac{L}{L_0}}^\beta\;,
}
with the matter power spectrum $P_{\delta}$. The best-fit parameters of this model are $A=5.76^{+0.60}_{-0.62}$, $\eta_{\rm other}=-0.27^{+0.80}_{-0.79}$, and $\beta=1.13^{+0.25}_{-0.20}$ (all marginalised and $1\sigma$). We adopt the pivot redshift $z_0=0.3$ and luminosity $L_0$, corresponding to an absolute magnitude of $-22$ in the $r$-band, from that work. 

The constant $C_1 = 0.0134/\rho_{\rm cr}$ is set for convenience to quantify the amplitude dependence in terms of the dimensionless parameter $A$, where $A=1$ corresponds to the normalisation determined by \citet{hirata04} using SuperCOSMOS data \citep{brown02}. The minus sign in Equation (\ref{eq:IAmodel}) accounts for the fact that the major axis of a galaxy is aligned with the \lq stretching\rq\ direction of tidal forces \citep[see][]{hirata04}.

The growth factor of matter fluctuations $D(z)$ and the matter power spectrum are evaluated for the Millennium cosmology, using the \citet{eisenstein98} transfer function and non-linear corrections by \citet{smith03}. The latter has been shown to fit the simulation matter power spectrum reasonably well on the relatively large scales we are interested in \citep[e.g.][]{hilbert09}, but note that potential amplitude mismatches would be calibrated via the galaxy bias measurements described below.

We now define galaxy samples from the simulation catalogues by selecting all galaxies with $z<0.7$ and $M_r<-19$ by default, so that their redshifts and luminosities are covered by the observational data sets. From a rest-frame colour-magnitude diagram of these galaxies, shown in the right panel of Fig.$\,$\ref{fig:cmd}, right panel, we determine a cut along the line $-7.2\,(M_u-M_r)-12.9$ which isolates the (bright part of the) red sequence and thus predominantly elliptical (and lenticular) galaxies. The sample consists of $23\,\%$ central early-type galaxies, $21\,\%$ central late-type galaxies (including lenticulars), as well as $16\,\%$ early-type and $39\,\%$ late-type satellites.

We compute the redshift and luminosity distributions of these sets of simulated galaxies and feed them into Equation (\ref{eq:IAmodel}), from which a model intrinsic alignment power spectrum is computed. The projected correlation function between the matter distribution and the radial intrinsic shear is then determined via \citep{hirata04}
\eqa{\nn
w_{\delta +}(r_p) &\hspace*{-0.3cm}=&\hspace*{-0.3cm} - \int \dd z\; {\cal W}(z) \int_0^\infty \frac{\dd k_\perp\, k_\perp}{2\,\pi}\; J_2(k_\perp r_p)\; P_{\delta {\rm I}} (k_\perp,z)\,,\\[-2ex]
\label{eq:wdeltaispec}
&&
}
where the kernel $J_\mu$ denotes the Bessel function of the first kind of order $\mu$. Here, ${\cal W}(z)$ is a weighting with the redshift distribution that takes the effects in a magnitude-limited sample into account (see the appendix of \citealp{mandelbaum11} for the explicit expression). In addition, we incorporate into ${\cal W}$ a linear weighting with the number of available galaxy pairs in a given transverse separation bin, which affects $w_{\delta +}(r_p)$ at the largest scales because pairs with large $r_p$ can only be found at higher redshift, due to the small mock survey size.

\begin{table}
\centering
\caption{Galaxy bias $b_{\rm g}$ in the simulation-based early-type galaxy samples. The full sample was used to produce Fig.$\,$\ref{fig:megaz_amplitude}, with cuts on redshift $z<0.7$ and rest-frame absolute $r$-band magnitude $M_r < -19$. The other samples correspond to the data points shown in Fig.$\,$\ref{fig:megaz_Lzdep}. Errors on $b_{\rm g}$ are $1\sigma$ and marginalised over $C$ as given in Equation (\ref{eq:wgg}). The third column lists the reduced $\chi^2$ of the fit.}
\begin{tabular}[t]{rcc}
\hline\hline
sample & $b_{\rm g}$ & $\chi^2_{\rm red}$\\
\hline\\[-2ex]
full sample & $1.66^{+0.03}_{-0.03}$ & 1.82\\
$0<z<0.5$ & $1.60^{+0.04}_{-0.04}$ & 0.52 \\
$0.5<z<0.7$ & $1.64^{+0.04}_{-0.04}$ & 1.17 \\
$0.7<z<0.9$ & $1.73^{+0.04}_{-0.04}$ & 2.00 \\
$0.9<z<1.2$ & $1.95^{+0.03}_{-0.03}$ & 1.97 \\
$-23.5<M_r<-20.5$ & $1.70^{+0.04}_{-0.03}$ & 0.80 \\
$-20.5<M_r<-19.0$ & $1.65^{+0.03}_{-0.03}$ & 1.05 \\
$-19.0<M_r<-18.3$ & $1.69^{+0.04}_{-0.04}$ & 0.56 \\
$-18.3<M_r<-17.0$ & $1.64^{+0.03}_{-0.04}$ & 1.12 \\[0.1ex]
\hline
\end{tabular}
\label{tab:galaxybias}
\end{table}

To obtain the corresponding signal from the simulation-based models, we calculate the correlation function 
\eq{
\label{eq:corr3D}
\xi_{\rm g +}(r_p,\Pi) = \frac{-1}{N_{\rm gal} (N_{\rm gal}-1)}\; \sum_{i \neq j}^{N_{\rm gal}} \epsilon_{+, i \rightarrow j}\; \Delta(r_p,\Pi)\;,
}
as a function of comoving transverse galaxy separation $r_p$ and line-of-sight separation $\Pi$. Here, $N_{\rm gal}$ is the number of galaxies in the sample and  $\epsilon_{+, i \rightarrow j}$ the galaxy ellipticity component of galaxy $i$ tangential to the line connecting galaxies $i$ and $j$. The function $\Delta(r_p,\Pi)$ is unity if galaxies $i$ and $j$ have separations that lie within the bins defined by $r_p$ and $\Pi$, and zero otherwise. The minus sign in Equation (\ref{eq:corr3D}) implies that radial alignment induces a positive correlation, as is customary in intrinsic alignment measurements. We achieve higher signal-to-noise (S/N) by summing the correlation function along the line of sight,
\eq{
\label{eq:corrprojected}
w_{\rm g+}(r_p) = \sum_{i=1}^{N_\Pi} \xi_{\rm g +}(r_p,\Pi_i) \Delta_\Pi\;,
}
where the summation runs over $N_\Pi$ bins of width $\Delta_\Pi$. We place $N_\Pi=18$ bins in the range $\Pi= \pm 90\,{\rm Mpc}/h$.

To convert $w_{\rm g +}$ to $w_{\delta +}$, we proceed analogously to the observational analyses, i.e. we assume $w_{\rm g +} = b_{\rm g} w_{\delta +}$ (linear and deterministic galaxy bias) and measure the clustering signal of the same galaxy sample in the linear and quasi-linear regime to determine $b_{\rm g}$. For convenience the galaxy clustering correlation function $w_{\rm gg}$ is measured as a function of angular separation $\theta$ rather than $r_p$ with the publicly available tree code \texttt{ATHENA}\footnote{\texttt{http://www2.iap.fr/users/kilbinge/athena}}, employing the estimator by \citet{landy93}.

The clustering correlation function is modelled as \citep{hirata07}
\eqa{
\label{eq:wgg}
w_{\rm gg}(\theta) &=& b_{\rm g}^2 \int \dd z\; {\cal W}(z)\\\nn
&& \hspace*{0.1cm} \times\; \int_0^\infty \frac{\dd k_\perp\, k_\perp}{2\,\pi}\; J_0\bb{k_\perp \theta\, \chi(z)}\; P_\delta (k_\perp,z) + C\;,
}
where $\chi(z)$ denotes comoving distance. The same matter power spectrum as used for evaluating Equation (\ref{eq:IAmodel}) is employed. The free parameters in Equation (\ref{eq:wgg}) are the linear bias $b_{\rm g}$ and the offset $C$ (to account for the undetermined integral constraint; see e.g. \citealp{joachimi11} for the analogous procedure), which are jointly fit to $w_{\rm gg}$ as determined from the simulation. To avoid the deeply non-linear clustering regime and effects of the survey aperture on large scales, the fit range is limited to $15^\prime < \theta < 75^\prime$, using 7 logarithmically spaced bins. 

Table~\ref{tab:galaxybias} summarises the resulting values of the galaxy bias in the different simulation-based samples that we consider. We reproduce the general trend of increased galaxy bias with higher redshift and higher luminosity (the latter only weakly though) of the galaxy sample \citep[cf.][]{guo12}. The measured $w_{\rm g +}$ are divided by $b_{\rm g}$, with the uncertainty on the bias included in the error bars.

\subsection{Intrinsic alignment amplitudes}
\label{sec:megazamplitude}

\begin{figure}
\centering
\includegraphics[scale=.36,angle=270]{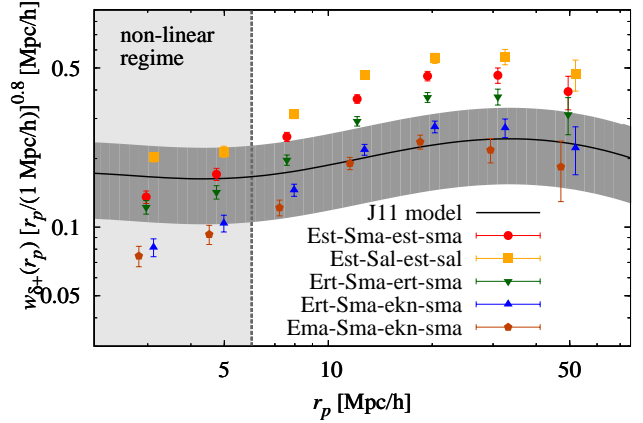}
\caption{Projected matter-galaxy ellipticity correlation function $w_{\delta +}$ as a function of transverse separation $r_p$. The black solid line is the J11 model obtained from the best-fit parameters in \citet{joachimi11}, i.e. the model that jointly fits the signal from the SDSS early-type galaxy samples considered in that work. The dark grey band is an estimate of the $1\sigma$ uncertainty of this model. Points with error bars correspond to the signals of the simulation-based galaxy models. Red circles (orange squares; green downward triangles; blue upward triangles; brown pentagons) correspond to the \texttt{Est-Sma-est-sma} (\texttt{Est-Sal-est-sal}; \texttt{Ert-Sma-ert-sma}; \texttt{Ert-Sma-ekn-sma}; \texttt{Ema-Sma-ekn-sma}) model. The light grey region indicates the regime with non-linear bias which hinders the conversion from $w_{\rm g+}$ to $w_{\delta +}$. Note that the ordinate axis has been rescaled by $[r_p/(1\,{\rm Mpc}/h)]^{0.8}$.}
\label{fig:megaz_amplitude}
\end{figure} 

In Fig.$\,$\ref{fig:megaz_amplitude} the resulting observation-based model (J11 model henceforth) for $w_{\delta +}(r_p)$ is plotted as the solid curve. On scales $r_p<6\,{\rm Mpc}/h$ the assumption of linear galaxy bias employed in the transition from galaxy-shape correlation to matter-shape correlation of the SDSS samples breaks down, so that the model is not reliable in this regime. The uncertainty of the model is estimated by propagating the $1\sigma$ errors on the fit parameters $A$, $\eta_{\rm other}$, and $\beta$, assuming that they are uncorrelated, which is a fair (see Fig.$\,$14 of \citealp{joachimi11}) and conservative assumption.

The resulting correlation functions $w_{\delta +}$ for several galaxy shape models are also shown in Fig.$\,$\ref{fig:megaz_amplitude}. Both the J11 model and the simulation signals agree fairly well in the dependence on $r_p$, but not in amplitude in all cases. For a more quantitative interpretation of Fig.$\,$\ref{fig:megaz_amplitude} we model the simulation-based signals $w_{\delta +}^{\rm sim}$ as a rescaled version of the fiducial J11 model, i.e. $w_{\delta +}^{\rm sim}(r_p) = {\cal N}\, w_{\delta +}^{\rm J11}(r_p)$, and fit the amplitude ${\cal N}$ in the range $10\,{\rm Mpc}/h < r_p < 60\,{\rm Mpc}/h$ (including the full covariance between $r_p$ bins and the contributions to the diagonal due to the errors on $b_{\rm g}$). 

Results for the four models under consideration are shown in Table~\ref{tab:megazamp}. As the reduced $\chi^2$ around unity of the fits suggest, the $r_p$ dependence of the simulation-based signals is indeed in good agreement with the J11 model, and therefore with the prediction of the linear alignment model, as well as the observations of the SDSS early-type samples. If we include the points at $r_p \approx 7\,{\rm Mpc}/h$ in the fit, the goodness of fit degrades substantially in all cases. Provided that our assumptions about linear biasing (as well as about the simple amplitude rescaling due to different values of $\sigma_8$) still hold in this regime, the discrepancy could point at deficiencies in the modelling of satellites and small-scale effects.

Only the model \texttt{Ema-Sma-ekn-sma} is consistent at $1\sigma$ with the fiducial J11 model, i.e. ${\cal N}=1$. All other combinations that we explore have amplitudes that are more than $2\sigma$ above the J11 model, where the most discrepant ones are based on early-type modelling with the simple inertia tensor. Switching to the reduced inertia tensor lowers the correlation function amplitude by $20\,\%$ (compare \texttt{Est-Sma-est-sma} to \texttt{Ert-Sma-ert-sma}), whereas the introduction of misaligned halo major axes leads to a factor 2 reduction (compare \texttt{Est-Sma-est-sma} to \texttt{Ema-Sma-ekn-sma}), in good agreement with \citet{okumura09}. The shape modelling of late types has significant impact on the analysis of these early-type samples, with a $20\,\%$ effect on overall amplitude (compare \texttt{Est-Sal-est-sal} to \texttt{Est-Sma-est-sma}). The inclusion of satellites is equally important, as can be seen from the $25\,\%$ amplitude reduction between models \texttt{ert} and \texttt{ekn}.

\begin{table}
\centering
\caption{Best-fit amplitude rescaling ${\cal N}$ of the J11 model to fit simulation results in the range $10\,{\rm Mpc}/h < r_p < 60\,{\rm Mpc}/h$. We also list $1\sigma$ errors and the reduced $\chi^2$ of the fit. Note that the uncertainty of the J11 model allows a rescaling between 0.63 and 1.37 at $1\sigma$ (corresponding to the dark grey region in Fig.$\,$\ref{fig:megaz_amplitude}).}
\begin{tabular}[t]{rcc}
\hline\hline
model & ${\cal N}$ & $\chi^2_{\rm red}$ \\
\hline\\[-2ex]
\texttt{Est-Sal-est-sal} & $2.37^{+0.09}_{-0.09}$ & 0.52 \\
\texttt{Est-Sma-est-sma} & $1.90^{+0.07}_{-0.07}$ & 1.38 \\
\texttt{Ert-Sma-ert-sma} & $1.53^{+0.06}_{-0.06}$ & 1.45 \\
\texttt{Ert-Sma-ekn-sma} & $1.13^{+0.05}_{-0.05}$ & 1.00 \\
\texttt{Ema-Sma-ekn-sma} & $0.97^{+0.05}_{-0.05}$ & 0.85 \\
\hline
\end{tabular}
\label{tab:megazamp}
\end{table}

\subsection{Redshift \& luminosity dependence}
\label{sec:megazlzdep}

To compare the redshift and luminosity dependencies of $w_{\delta +}$ for the different models, we split the default mock galaxy sample ($z<0.7$; $M_r < -19$) at $z=0.5$ into two redshift bins with a similar number of galaxies. Then we extend the sample by two extra redshift bins with boundaries at $z=0.9$ and $z=1.2$, keeping the constraint $M_r<-19$. Likewise, we divide the original full sample into two magnitude bins at $M_r=-20.5$ and add two more magnitude bins with boundaries at $M_r=-18.3$ and $M_r=-17$, retaining the original redshift cut $z<0.7$.

\begin{figure}
\centering
\includegraphics[scale=.34,angle=270]{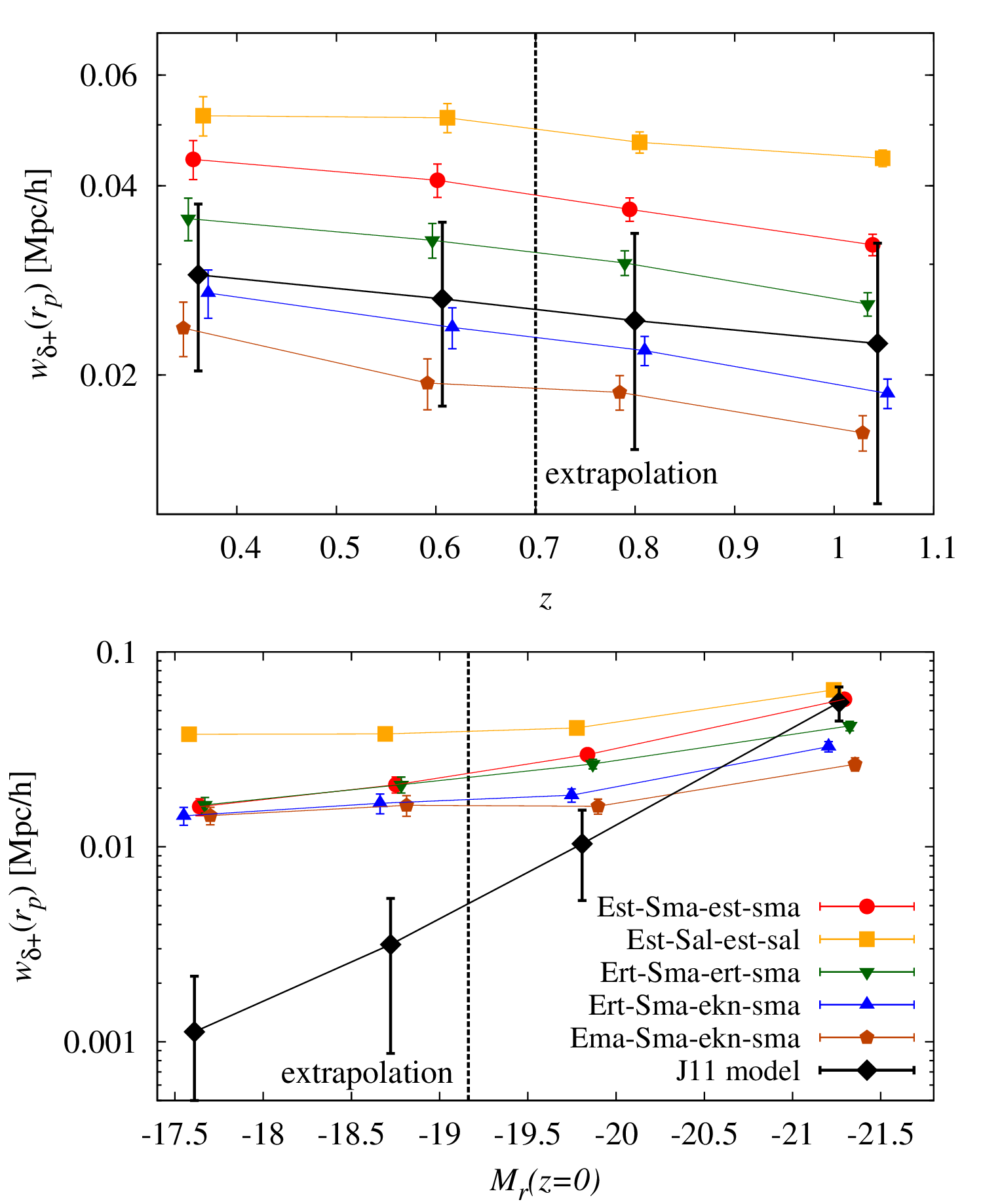}
\caption{\textit{Top panel}: Redshift dependence of $w_{\delta +}$. Transverse separation scales have been averaged over in the (quasi-)linear regime $6< r_P/[Mpc/{\rm h}]<35$. Black diamonds correspond to the J11 model obtained from the best-fit parameters in \citet{joachimi11}; error bars are propagated from the errors on these parameters. Symbols and colours of the simulation-based models are the same as in Fig.$\,$\ref{fig:megaz_amplitude}. Note that the two low-redshift bins cover the range for which observational data was available to determine the J11 model. \textit{Bottom panel}: Same as above, but showing the rest-frame absolute $r$-band magnitude $M_r$ dependence of $w_{\delta {\rm g+}}$. Note that the two brightest magnitude bins cover the range for which observational data was available to determine the J11 model.}
\label{fig:megaz_Lzdep}
\end{figure}

Note that the additional bins reach substantially beyond the ranges probed by the observations in \citet{joachimi11}, so that the observation-based model has to be extrapolated with increasing uncertainty. For each redshift and magnitude bin we proceed as outlined in Section$~$\ref{sec:megazmethod} to calculate $w_{\delta +}$ in a single broad $r_p$ bin that covers the (quasi-)linear range between $6\,{\rm Mpc}/h$ and $35\,{\rm Mpc}/h$. The resulting signals are given in Fig.$\,$\ref{fig:megaz_Lzdep}, again including the error on $b_{\rm g}$ into the error bars of the simulation-based models, and propagating the error of the fit parameters in Equation (\ref{eq:IAmodel}) into the uncertainty on the J11 model.

All signals agree well in an overall weak decrease of the amplitude of $w_{\delta +}$ with redshift, including the redshift range $0.7 < z< 1.2$ in which the J11 model is not directly constrained by observations. In stark contrast, the J11 and simulation-based models are clearly discrepant in the dependence on absolute magnitude. Observations, in particular of the SDSS LRG samples, clearly favour a significant increase of intrinsic alignments with higher galaxy luminosity, which the J11 model extrapolates to $M_r \ga -19$. 

The simulation-based models predict a shallower slope at the bright end and weak or no dependence on absolute magnitude for fainter galaxies. This breakdown could either indicate that the roughly linear luminosity dependence observed for the SDSS samples does not extend to fainter galaxies, or that our shape models for faint galaxies, mostly satellites (the satellite fractions in the samples corresponding to the four bins shown in Fig.$\,$\ref{fig:megaz_Lzdep}, bottom panel, are $44\,\%$, $67\,\%$, $88\,\%$, and $95\,\%$, from bright to faint), produce too strong alignment. Both scenarios are plausible and can be tested with forthcoming improved observations and simulations.

This latter result limits the validity of predictions of intrinsic alignment contamination made with the simulation-based models. However, note that that a typical galaxy in a \textit{Euclid}-like survey has an absolute rest-frame magnitude of $M_r \approx -20$ around the median redshift, i.e. it lies in a range where the simulation performs reasonably well. The intrinsic alignment signal of faint galaxies at low redshift is likely to be over-predicted though, in particular on small scales where satellite galaxies contribute most.

\subsection{Late-type observations}
\label{sec:wigglezmethod}

We also attempt a comparison with the intrinsic alignment measurement by \citet{mandelbaum11}, the only analysis to date that includes a late-type galaxy sample covering a substantial range in redshift. It is based on the WiggleZ Dark Energy Survey \citep{drinkwater10} which is a spectroscopic redshift survey of bright emission-line galaxies that obtained about $200,000$ redshifts over a total of about $1000\, {\rm deg}^2$ of equatorial sky out to $z \la 1$. The northern parts of the survey overlap with SDSS imaging from which \citet{mandelbaum06} determined more than $10^6$ shapes of galaxies (see that work for details about the shape measurements). The overlap region contains $76,000$ galaxies with reliable redshifts from WiggleZ of which about a third have high-quality shape measurements. The sample is additionally split into two redshift bins at $z=0.52$.

The criteria for the WiggleZ target selection rely on UV photometry which we do not have at our disposal in the Millennium galaxy catalogues. Besides, this would again involve the direct application of colour cuts. Instead, we impose the constraint $20 < r < 21.8$, where the faint limit originates from shape measurement catalogue while the bright limit is an original WiggleZ cut to remove contamination by low-redshift objects, to construct another colour-magnitude diagram (see Fig.$\,$\ref{fig:cmd}, left panel). As Fig.$\,$3 in \citet{wyder07} shows, the WiggleZ colour cut ${\rm NUV}-r<2$ selects roughly the bluest third of the blue cloud. We reproduce this selection approximately by introducing a cut $M_u-M_r<0.35$.

This is supplemented by the criterion $M_r < -3.53\, (M_u-M_r) - 20.5$, which avoids a spurious contamination with very blue, very bright galaxies classified as early types; see Fig.$\,$\ref{fig:cmd}. It is known that the \citet{bower06} models overpredict the abundance of luminous blue galaxies, and that the bulge-to-total luminosity ratios of the full galaxy sample tend to large values for $M_r \la -21$ \citep{gonzalez09}. Therefore we conclude that these galaxies are not representative of galaxies in WiggleZ and exclude them from further analysis.

We obtain $w_{\rm g +}$ as well as the correlation function $w_{++}(r_p)$ which is calculated by analogy to Equations (\ref{eq:corr3D}) and (\ref{eq:corrprojected}), except that in this case the tangential ellipticities of galaxy pairs are correlated. As many WiggleZ galaxies are at redshifts $z \ga 0.5$, this quantity could be contaminated by correlations due to gravitational shear. However, we find that including the contribution to galaxy ellipticity by shear leaves $w_{++}$ practically unchanged, in line with the conclusions of \citet{mandelbaum11}.

We find that all simulation-based correlations are consistent with zero over the same scales as shown in Fig.$\,$3 of \citet{mandelbaum11}, with very little dependence on the model assumptions. The associated $p$-values (the probability of obtaining a $\chi^2$ value at least as extreme as the one observed assuming that the hypothesis of a zero signal is true) generally exceed 0.5; only for $w_{\rm g +}$ in the high-redshift bin $p \approx 0.15$. This is in line with the null detection of intrinsic alignments in the WiggleZ data, but note that error bars are more than an order of magnitude larger over all scales in the observational data sets, hence there is little discriminatory power.

A caveat in this direct comparison of $w_{\rm g +}$ measurements in WiggleZ and the Millennium Simulation is that we have assumed that our models do not only provide a fair description of the correlation between galaxy shape and the matter distribution but also of the link between matter and the galaxy distribution. In other words, we have implicitly assumed that the simulated galaxies have the same galaxy bias as the observed WiggleZ galaxies, which is not guaranteed because of the different sample selection criteria and the high $\sigma_8$ value of the simulation. However, the modelled and observed results would still be broadly consistent even with a difference in galaxy bias by a factor of two or more.

The extremely blue WiggleZ galaxies are rare objects, which leads to the large error bars on correlation functions due to low number densities, and exhibit very little intrinsic alignment, as confirmed by our results. This makes them unfavourable for intrinsic alignment investigations. A large-area survey to similar depths of more generic late-type galaxies, which also make up the bulk of samples used for weak lensing studies, would be desirable but is not available to date.

At lower redshift the blue subsamples of SDSS spectroscopic galaxies studied by \citet{hirata07} could potentially provide another test of our models, but applying the corresponding selection criteria to the simulated catalogues produces only a few hundred galaxies per line of sight, so that any results would be inconclusive due to large error bars. A selection based on a colour-magnitude diagram as for the WiggleZ samples is problematic as \citet{hirata07} split the population by observed rather than rest-frame magnitude which leads to an uncertain yet significant contamination of the blue sample by early-type galaxies (see the discussions in \citealp{hirata07} and \citealp{joachimi11}). Hence we refrain from a detailed analysis.

\section{Impact on weak gravitational lensing measurements}
\label{sec:impact}

One of the main motivations for studying intrinsic shapes and alignments of galaxies, beside the insights into galaxy formation and evolution, is to assess the contamination of weak gravitational lensing measurements. The ability to extract information on the growth of structure and the geometry of the Universe from the two-point function of gravitational shear (GG henceforth) may be severely limited by correlations between the intrinsic shapes of galaxies (II henceforth) as well as correlations between intrinsic shapes of foreground galaxies and the shear acting on background galaxy images (GI henceforth). In the following we will assess the impact of the galaxy shape model that fared best in the comparison to observational data sets on a cosmic shear measurement with typical parameters of forthcoming surveys.

\subsection{Survey design}
\label{sec:iacontsurveys}

Combining the results of Section$~$\ref{sec:iaobs} with those of Paper I, the model \texttt{Ert-Sma-ekn-sma} provides overall fair agreement with observational constraints and will hence be employed for the intrinsic alignment predictions of this section. The constraints on late-type intrinsic alignments and shape distributions are still weak or inconclusive. We take the observed suppression of the simulated correlations for the \lq red\rq\ SDSS early-type galaxies due to the choice of late-type model (see Figs.$\,$\ref{fig:megaz_amplitude} and \ref{fig:megaz_Lzdep}) as an indication in favour of the \texttt{Sma;sma} models which are expected to yield the smallest signal (see also Section$~$\ref{sec:modeleffect}).

The model combination \texttt{Ert; ekn} produces the best agreement in amplitude and redshift dependence with SDSS early-type observations. It also fares well at describing the intrinsic ellipticity distribution of a typical early-type galaxy sample in COSMOS (see Figs.$\,$5 and 7 in Paper I). Moreover, the choice \texttt{Ert-Sma-ekn-sma} should be slightly more conservative in terms of the expected weak lensing contamination than \texttt{Ema-Sma-ekn-sma}, the only other model combination tested that provides a good match in amplitude, as the resulting intrinsic alignment signal is larger at all redshifts (see Fig.$\,$\ref{fig:megaz_Lzdep}, top panel).

\begin{table}
\centering
\caption{Properties of the galaxy samples used in Section$~$\ref{sec:impact}. Given are the passband and magnitude limit that define the sample, as well as the expected median redshift $z_{\rm med}$ of the corresponding surveys.}
\begin{tabular}[t]{lccc}
\hline\hline
survey & band & mag. limit & $z_{\rm med}$\\
\hline\\[-2ex]
\textit{Euclid}-like & $RIZ$ & 24.5  & 0.9 \\
KiDS-like   & $r$   & 24.4  & 0.7 \\
KiDS-like   & $g$   & 24.6  & 0.7 \\
shallow     & $r$   & 23.9  & 0.6 \\
\hline
\end{tabular}
\label{tab:surveydepths}
\end{table}

By default we work with a galaxy sample similar to the one anticipated for the \textit{Euclid} wide imaging survey, by requiring a limiting AB magnitude of 24.5 (for $10\sigma$ extended sources) in the broad $RIZ$ \textit{Euclid} passband \citep{laureijs11}. As magnitudes in this filter are not available from the semi-analytic models, we approximate $RIZ$ fluxes, which cover the wavelength range $550-900\,$nm, by adding the fluxes of the SDSS $r$ and $i$ filters, as well as half of the $z$ filter (whose throughput lies between $800-1000\,$nm). These assumptions result in a total sample with a median redshift of $z_{\rm med}=0.9$ and a mean galaxy number density of $n_{\rm g}=28.5\,{\rm arcmin}^{-2}$, which is in good agreement with the numbers quoted in \citet{laureijs11}, $z_{\rm med} \ga 0.9$ and $n_{\rm g}=30\,{\rm arcmin}^{-2}$. 

We will also consider a survey design closer to forthcoming ground-based surveys such as KiDS \citep{jong13}. Adopting the magnitude limits given in \citet{laureijs11}, we obtain a mock sample with a depth of $z_{\rm med}=0.7$, in good agreement with other predictions for weak lensing science. We construct two KiDS-like samples selected in the $g$ and $r$ bands, respectively, as well as a shallower survey (0.5 magnitudes brighter limit, $z_{\rm med}=0.6$) for comparison. The corresponding properties of the galaxy samples are summarised in Table~\ref{tab:surveydepths}.

All current and future cosmic shear surveys will make use of photometric redshift information to study the large-scale structure in tomographic slices. We do not take into account the scatter along the line of sight introduced by photometric redshift estimation, but defer analysing the impact of photometric redshift scatter on the intrinsic alignment signals and their control to future work. This implies in particular that the II signal is restricted to redshift auto-correlations in our study, and that these auto-correlations are expected to be weaker in reality due to photometric redshift scatter.

\subsection{Method}
\label{sec:iacontmethod}

The correlations induced by weak lensing are generated along the line of sight and thus best measured as a function of angular separation, in contrast to intrinsic alignments which depend on the physical separation of the galaxies correlated. Therefore we now use the standard correlation functions \citep[e.g.][]{Schneider02}
\eq{
\label{eq:xipm}
\xi_{\pm}(\theta) = \ba{ \epsilon_+(\vek{\vartheta})\, \epsilon_+(\vek{\vartheta}+\vek{\theta}) }_{\svek{\vartheta}} \pm  \ba{ \epsilon_\times(\vek{\vartheta})\, \epsilon_\times(\vek{\vartheta}+\vek{\theta}) }_{\svek{\vartheta}}\;,
}
where $\ba{\cdot}_{\svek{\vartheta}}$ denotes the average over all angular separation vectors $\vartheta$\footnote{Three types of correlation functions are used in this work: the angular correlation functions $\xi_{\pm}$, the correlation function $\eta(r)$ (see Equation \ref{eq:3Dcorr}) which depends on physical separation $r$ but is otherwise identical, and the correlation functions $w_{++}(r_p)$ and $w_{\rm g +}(r_p)$ (as employed in Section$~$\ref{sec:iaobs}) which depend on transverse separation $r_p$ and additionally stack the signal from several line-of-sight bins.}. Note that in this and the following equations we suppress indices indicating the correlation of redshift bin subsamples for notational convenience. The correlation functions $\xi_{\pm}$ are the measures of choice to be applied to galaxy catalogues derived from real data as they are not affected by masking or a complex survey geometry.

As gravitational lensing creates only curl-free shear patterns to first order, it is desirable to separate the two-point statistics of the shear field into a curl-free E-mode and a divergence free B-mode. This can be achieved by means of the aperture masses \citep{Schneider98}
\eq{
\label{eq:mapdef}
M_{\rm E,B}(\theta) = \int \dd^2 \vartheta\; Q(|\vek{\vartheta}|)\; \epsilon_{+,\times}(\vek{\vartheta})\;,
}
where $Q$ is an arbitrary axi-symmetric weight function. The corresponding two-point statistics, the aperture mass dispersions, can be determined from the correlation functions via\footnote{For the simple square geometry in the simulations and shears/ellipticities on a grid the aperture mass dispersion could be directly computed via Equation (\ref{eq:mapdef}); see e.g. \citet{hilbert09}. As we are interested in the relevance of intrinsic alignments for a realistic mock survey and data analysis, we opt for an approach that could readily be applied to real data.}
\eqa{\nn
\ba{M_{\rm E,B}^2}(\theta) &\hspace*{-0.3cm}=&\hspace*{-0.3cm} \frac{1}{2}\!\! \int_0^{2\theta} \!\!\frac{\dd \vartheta\, \vartheta}{\theta^2}\! \bb{\xi_+(\vartheta)\, T_+\br{\frac{\vartheta}{\theta}} \pm \xi_-(\vartheta)\, T_-\br{\frac{\vartheta}{\theta}}},\\
\label{eq:map}
&&
}
where $T_\pm$ are weight functions whose form is explicitly given in \citet{SvWM02} for the simplest polynomial weight function $Q$ introduced by \citet{Schneider98}, which we also employ.

In practice the integration of Equation (\ref{eq:map}) is executed as a Riemann sum over very finely binned correlation functions ($3000$ logarithmically spaced angular separation bins). Errors on $\xi_{\pm}$ are propagated into $\ba{M_{\rm E,B}^2}$ by the formalism described in \citet{pen02} and \citet{heymans05}. The smallest $\theta$ bin entering the Riemann sum is given by the minimum separation for which $\xi_{\pm}$ is measured. As this is always larger than zero, some leakage from E-modes to B-modes and vice versa is expected\footnote{This leakage can in principle be reduced by extrapolating signals rather than setting them to zero below the minimum separation. However, the intrinsic alignment correlation functions are very noisy and a functional form of the angular dependence at these small scales is unknown, prompting us to avoid extrapolation in this case.}. Consulting the results of \citet{kilbinger06}, we choose the minimum angular separation such that any leakage has a negligible effect on our E-mode signals ($0.6\,^{\prime\prime}$). The aperture-mass dispersions are calculated for 12 logarithmically spaced angular bins in the range $1-100\,^\prime$.

It is convenient to express the total intrinsic alignment contribution relative to the weak lensing signal at a given scale via
\eq{
\label{eq:riatheta}
r_{\rm IA}(\theta) \equiv \frac{|\ba{M_{\rm E,II}^2}(\theta)+\ba{M_{\rm E,GI}^2}(\theta)|}{\ba{M_{\rm E,GG}^2}(\theta)}\;,
}
where, by using the absolute value in the numerator, we have accounted for the fact that the GI signal is typically negative \citep[e.g.][]{hirata04}. Errors on $r_{\rm IA}$ are propagated from the full covariances of $\ba{M_{\rm E,II}^2}$, $\ba{M_{\rm E,GI}^2}$, and $\ba{M_{\rm E,GG}^2}$, neglecting any cross-variances between the three statistics.

To further condense this information, we seek to devise a single ratio $R_{\rm IA}$ of intrinsic alignment over lensing signal for a given redshift bin combination. Since a logarithmic angular binning for the aperture-mass dispersion is used, we deem a simple average over angular bins appropriate, if statistical error and cross-correlation between these bins are taken into account. Hence, introducing the shorthand notation $r_{{\rm IA},i} \equiv r_{\rm IA}(\theta_i)$ and ${\cal C}_{ij} \equiv {\rm Cov}\br{r_{{\rm IA},i};r_{{\rm IA},j}}$, we define the covariance-weighted average
\eqa{
\label{eq:ria}
R_{\rm IA} &\!\!\!\equiv&\!\!\! \sigma_R^2 \sum_{i,j=b_{\rm min}}^{b_{\rm max}} \bc{{\cal C}^{-1}}_{ij} r_{{\rm IA},j}\;;\\ \nn
\sigma_R &\!\!\!=&\!\!\! \br{ \sum_{i,j=b_{\rm min}}^{b_{\rm max}}  \bc{{\cal C}^{-1}}_{ij} }^{-1/2}\;.
}
Here, $\sigma_R$ denotes the standard deviation of $R_{\rm IA}$. We fix $b_{\rm max}=12$, i.e. at the highest separation bin, and set $b_{\rm min}=7$, corresponding to an angular range $10-100\,^\prime$ included in $R_{\rm IA}$. The lower limit of $10\,^\prime$ is motivated by the lack of observational constraints on the two-point statistics on smaller scales (at the median redshift of \textit{Euclid} a $10\,^\prime$ angular separation corresponds to $r_p \approx 6\,{\rm Mpc}/h$; compare this to Fig.$\,$\ref{fig:megaz_amplitude}). Moreover, the small-scale intrinsic alignment signal is strongly dominated by satellite galaxies for which our shape models are most uncertain.

\subsection{Angular dependence}
\label{sec:angdep}

\begin{figure}
\centering
\includegraphics[scale=.34,angle=270]{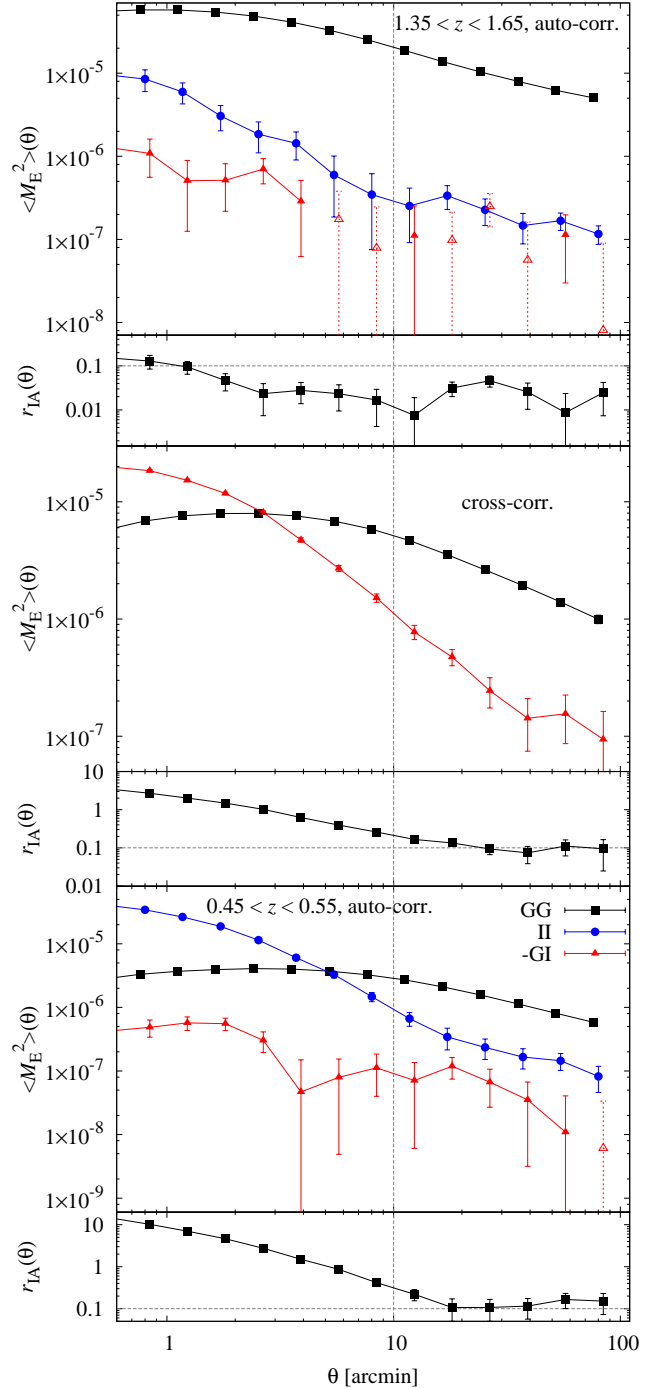}
\caption{Aperture-mass dispersion $\ba{M_{\rm E}^2}(\theta)$ for gravitational lensing (GG) and intrinsic alignment signals (GI and II) for three redshift bin combinations in a \textit{Euclid}-like survey (${\rm RIZ}<24.5$; $z_{\rm med} \approx 0.9$). The black squares correspond to the GG signal, the blue circles to the II signal, and red triangles to the negative GI signal (open symbols show the signal where positive). In the small panels the ratio of total intrinsic alignment over lensing signal $r_{\rm IA}(\theta)$ is plotted. Grey dotted lines are used to indicate an $r_{\rm IA}$ threshold of $10\,\%$, as well as $\theta=10\,^\prime$, which serves as a rough separator between the non-linear and (quasi-)linear regimes. \textit{Top panels}: Auto-correlations in a redshift bin defined by $1.35<z<1.65$. \textit{Centre panels}: Cross-correlations between two redshift bins with $1.35<z<1.65$ and $0.45<z<0.55$, respectively. \textit{Bottom panels}: Auto-correlations in a redshift bin defined by $0.45<z<0.55$.}
\label{fig:angular_example}
\end{figure} 

In Fig.$\,$\ref{fig:angular_example} the E-mode aperture mass dispersion of the GG, GI, and II signals is shown as a function of filter scale $\theta$. We use a galaxy sample representative of a \textit{Euclid}-like survey, presenting auto-correlations in a low-redshift bin ($0.45<z<0.55$) and a high-redshift bin ($1.35<z<1.65$), as well as cross-correlations between the two samples. Note that in particular the II signal strength depends on the redshift bin width. We have chosen the widths to be close to the minimum that is realistically possible for a \textit{Euclid}-like survey, so that the II correlations shown represent an upper limit. 

The cosmic shear signal from the auto-correlations is in good agreement with the results by \citet{hilbert09} who computed $\ba{M_{\rm E,GG}^2}(\theta)$ directly from the shear field at $z \approx 1$. The II signal is clearly detected on all scales considered in the auto-correlations at low and high redshift, increasing strongly for smaller angular separations. The GI signal is negative where significant, i.e. foreground ellipticities and the gravitational shear on the images of galaxies in the background are preferentially anti-correlated. It remains subdominant in amplitude to II in the auto-correlations (because the redshift bins are relatively narrow), but has a strong signal with a steep dependence on $\theta$ in the cross-correlation.

Figure \ref{fig:angular_example} also provides the ratio of the total intrinsic alignment correlations over the lensing signal, calculated via Equation (\ref{eq:riatheta}). Somewhat arbitrarily, we use $r_{\rm IA}(\theta)=0.1$ as a threshold above which the intrinsic alignment contamination is considered severe. Future surveys like \textit{Euclid} strive to measure cosmic shear two-point statistics to sub-percent accuracy, so that any intrinsic alignment effects at the per cent level and higher need to be modelled or removed. Any uncertainty in the knowledge about the intrinsic alignment signals propagates through to increased errors on cosmology; the stronger the signal the larger the impact on constraints. If these effects exceed $\sim 10\,\%$, the requirements on the accuracy of the intrinsic alignment models and on the knowledge of their free parameters will become increasingly strict and hence challenging to meet.

The \textit{Euclid} baseline restricts the cosmological weak lensing analysis to redshifts larger than 0.5, mainly because the lensing signal at low redshift is small while intrinsic alignments are comparatively strong. For the auto-correlation of the low-redshift bin, which is located at this transitional redshift, $r_{\rm IA}(\theta) \ga 10\,\%$ on scales $\theta \ga 10\,^\prime$, and $r_{\rm IA}(\theta) > 1$ below $5\,^\prime$. Thus, our models suggest that for $z \la 0.5$ intrinsic alignments become too strong to allow for a precise cosmological measurement via weak lensing, whereas at higher redshifts this is feasible if the intrinsic alignment signal is well understood and the analysis is not extended too far into the non-linear regime. On these small scales the exploitation of cosmic shear measurements remains limited because of the difficulties of modelling the non-linear contributions to the matter power spectrum, the coupling of modes due to non-linear evolution, and the impact of baryonic physics \citep[e.g.][]{kiessling11,semboloni11b}. 

At $z \sim 1.5$ the intrinsic alignment contamination is a few per cent on all relevant scales, which is still too high to be left unaccounted for. This is worrisome due to the absence of deep, large-area spectroscopic surveys for representative galaxy samples needed to study intrinsic alignments at these redshifts. However, we caution again that in practice some dilution of the II signals is expected because of photometric redshift scatter. This does not apply to the GI signal in the cross-correlation which yields $r_{\rm IA}(\theta) \approx 10\,\%$ in the well constrained regime above $\theta = 10\,^\prime$. Since the model-independent removal of the GI signal severely reduces the cosmological power of the weak lensing survey \citep{joachimi08b,joachimi09}, this result implies tight requirements on the quality of intrinsic alignment models over a wide range in redshift.

\subsection{Redshift dependence}
\label{sec:cszdep}

\begin{figure}
\centering
\includegraphics[scale=.34,angle=270]{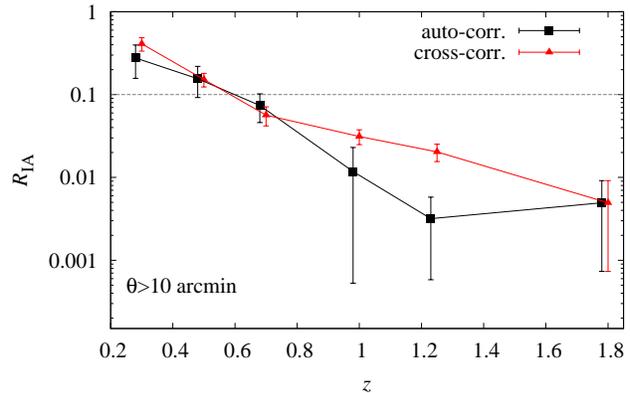}
\caption{Ratio of mean intrinsic alignment over lensing signals $R_{\rm IA}$ as a function of redshift for a \textit{Euclid}-like survey (${\rm RIZ}<24.5$; $z_{\rm med} \approx 0.9$), averaged over angular scales $10\,^\prime<\theta<100\,^\prime$. Auto-correlations of redshift bins as a function of the mean redshift in a bin are shown as black squares. Cross-correlations between a foreground redshift bin whose mean redshift is given on the abscissa and the background bin at the highest redshift ($1.4<z<2.2$) are shown as red triangles. The grey dotted line indicates an $R_{\rm IA}$ of $10\,\%$.}
\label{fig:angular_redshift}
\end{figure} 

To illustrate the redshift dependence of the intrinsic alignment contamination, we define six redshift bins containing roughly equal numbers of galaxies for a \textit{Euclid}-like survey, using the bin boundaries $\bc{0.2,0.4}$; $\bc{0.45,0.55}$; $\bc{0.65,0.75}$; $\bc{0.9,1.1}$; $\bc{1.1,1.4}$; $\bc{1.4,2.2}$. For each bin we calculate the averaged ratio $R_{\rm IA}$ (see Equation \ref{eq:ria}) for the auto-correlation, as well as the cross-correlation between each bin and the highest redshift bin. As the GI signal increases slightly stronger with redshift than the lensing signal for a given foreground redshift bin \citep[e.g.][]{joachimi10}, due to the different weights in the respective Limber equations, $R_{\rm IA}$ of the cross-correlation with the highest redshift bin in the background is largest, i.e. we consider the worst case. 

As is shown in Fig.$\,$\ref{fig:angular_redshift}, auto- and cross-correlations behave similarly in that $R_{\rm IA}$ decreases considerably with redshift, from close to 0.5 at $z \approx 0.3$ to a few per cent or less at $z \ga 1.2$. At lower redshift there is less structure along the line of sight that has contributed to the gravitational shear, causing a lower GG signal, whereas a given angular range probes smaller physical galaxy separations at which intrinsic alignment is stronger. This redshift range, below $z = 0.5$, is hence well suited to provide strong constraints on intrinsic alignments in a joint tomographic weak lensing analysis \citep[see][]{heymans13}. 

At $z \ga 0.6$ the shape model predicts an intrinsic alignment contamination below $10\,\%$ on average, so that the sensitivity of a cosmological weak lensing analysis to intrinsic alignments is somewhat reduced \citep[cf.][]{mandelbaum11,joachimi11}. The ratio $R_{\rm IA}$ for the GI signal remains significant at a few per cent at $z \ga 1$ where direct observational constraints on intrinsic alignments will be difficult to obtain, limiting the fidelity of any modelling attempt.

\subsection{Effect of galaxy sample selection}
\label{sec:depthfilter}

A reliable intrinsic alignment model that captures the dependence on galaxy luminosity, colour, and redshift accurately can in principle be used to optimise a weak lensing survey towards minimum contamination. For instance, one naively expects that a survey with a brighter apparent magnitude limit has higher intrinsic alignment contamination because at any given redshift galaxies will on average be more luminous and thus more strongly aligned.

\begin{figure}
\centering
\includegraphics[scale=.34,angle=270]{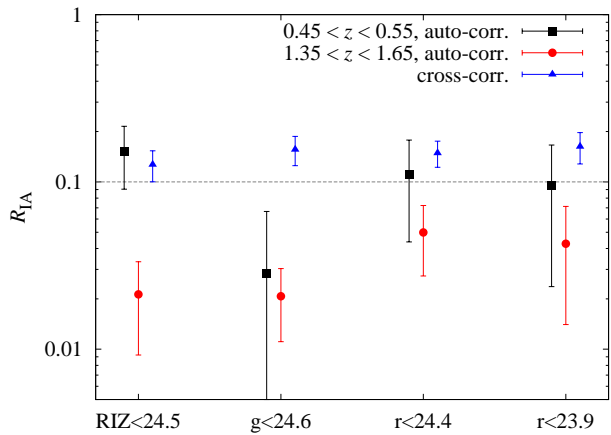}
\caption{Ratio of mean intrinsic alignment over lensing signals $R_{\rm IA}$, averaged over angular scales $10\,^\prime<\theta<100\,^\prime$, for four different combinations of survey parameters (passband and limiting magnitude; see Table~\ref{tab:surveydepths}). Black squares (red dots; blue triangles) correspond to auto-correlations in the redshift bin $0.45<z<0.55$ (auto-correlations in the redshift bin $1.35<z<1.65$; cross-correlations between the two bins). The grey dotted lines indicates $R_{\rm IA}=10\,\%$.}
\label{fig:angular_depth}
\end{figure} 

Generally, one seeks to optimise the statistical power of a survey by keeping the total exposure time fixed and trading off survey area and depth. Considering the weak lensing signal in isolation, this leads to maximising the area once a certain medium depth is reached \citep{amara07}. When including intrinsic alignments, this conclusion is reversed for two reasons: a) the relative strength of intrinsic alignments with respect to the cosmological signal becomes weaker if less luminous galaxies become part of the sample, and b) a deeper survey allows for a longer baseline in redshift, hence facilitating the calibration of intrinsic alignments with a minimum of assumptions about models \citep{joachimi10}.

As a precursory study, we investigate the dependence of $R_{\rm IA}$ on the four different survey setups listed in Table~\ref{tab:surveydepths}, varying the limiting magnitude and the passband. In Fig.$\,$\ref{fig:angular_depth} we have plotted $R_{\rm IA}$ for the two redshift auto-correlations and the cross-correlation of the same low- and high-redshift bins used in Section$~$\ref{sec:angdep}.

At low redshift brighter magnitude limits affect the galaxy population only marginally by removing the faintest objects, hence the small differences for the surveys observing in red filters. In the $g$-band survey $R_{\rm IA}$ is reduced (although the significance is low due to large error bars), which suggests that preferentially selecting blue, i.e. late-type, galaxies could be effective at reducing the intrinsic alignment contamination. Alternatively, a joint weak lensing analysis of two samples selected in different passbands could help to discern the lensing and intrinsic alignment signals by taking advantage of the achromatism of the lensing effect.

The intrinsic alignment contamination in the redshift cross-correlation is essentially unaffected by modifications of the survey parameters. As this signal is dominated by GI, this implies that the alignment of the different foreground samples with the surrounding large-scale structure that lenses the background sources is on average unchanged. At high redshift the impact of the magnitude limit is expected to be stronger. Indeed, $R_{\rm IA}$ for the \textit{Euclid}-like survey is lowest, but the significance is low, and switching between $r<24.4$ and $r<23.9$ does not show the same trend. Similar to the low-redshift auto-correlation, selecting galaxies in the $g$-band rather than $r$ slightly reduces $R_{\rm IA}$.

\begin{figure}
\centering
\includegraphics[scale=.34,angle=270]{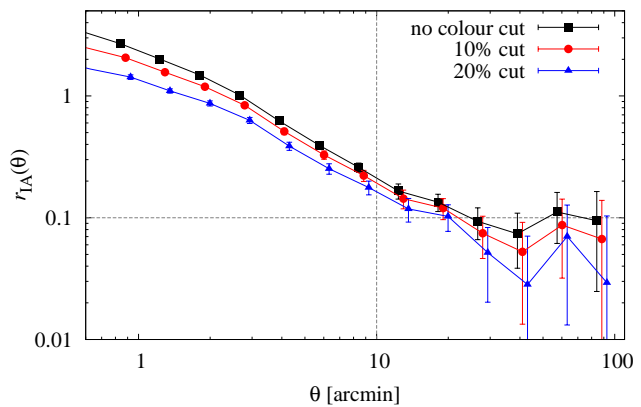}
\caption{Effect of colour cuts on the intrinsic alignment contamination. Shown is the ratio of GI over GG signal $r_{\rm IA}(\theta)$ for the cross-correlation between redshifts in the range $0.45<z<0.55$ and $1.35<z<1.65$. Black squares correspond to $r_{\rm IA}(\theta)$ without any cuts (for reference; same as in Fig.$\,$\ref{fig:angular_example}), red dots (blue triangles) to the signal when $10\,\%$ ($20\,\%$) of the reddest galaxies in the foreground sample have been removed using the cut $g-i>2.54$ ($g-i>2.42$). The ratio $r_{\rm IA}(\theta)$ is always computed using the lensing signal for the full sample to keep error bars small.}
\label{fig:angular_ccut}
\end{figure} 

The II signal is relatively straightforward to suppress by removing physically close galaxy pairs, as demonstrated by \citet{heymans06}. This procedure is not effective on the more worrisome GI contamination in cross-correlations of galaxy samples. Using the same setup as in Section$~$\ref{sec:angdep}, we test how well a simple cut in observed $g-i$ colour to remove the reddest galaxies in the low-redshift sample is capable of reducing the GI contamination. We consider two cases, cutting $10\,\%$ and $20\,\%$ of the sample via $g-i>2.54$ and $g-i>2.42$, respectively.

Figure$~$\ref{fig:angular_ccut} shows that increasingly strict colour cuts can suppress the GI signal on all scales. For the $20\,\%$ cut we find a reduction in signal by almost a factor of two around $1^\prime$ and for $\theta \ga 30^\prime$, and by about $25\,\%$ around $10^\prime$. This cut would imply an increase of the error bars on the cosmic shear signal by at most $12\,\%$ (on small scales where sample variance is negligible).

Cuts in rest-frame colour and absolute magnitude may prove to be even more effective in reducing intrinsic alignment contamination, but their performance depends strongly on the accuracy and precision of photometric redshift estimates. The same applies to the close-pair removal tailored towards the suppression of the II signal. These mitigation techniques will be assessed in detail in forthcoming work by combining the semi-analytic models presented here with mock photometric redshift catalogues.

\subsection{B-modes}
\label{sec:bmodes}

The gravitational deflection of light does not create B-modes in the resulting shear field to first order, so that two-point statistics of B-modes are routinely used to test for residual systematic effects. \citet{heymans06} detected B-modes generated by GI correlations in their simulations and proposed to use these as a diagnostic for intrinsic alignments.

We calculate the B-mode aperture mass dispersion $\ba{M_{\rm B}^2}(\theta)$ for similar redshift bin combinations as used in Section$~$\ref{sec:angdep}, again assuming a \textit{Euclid}-like survey. Both lensing and intrinsic alignment B-mode signals are generally much smaller than the E-modes ($|\ba{M_{\rm B}^2}(\theta)|<10^{-7}$ on scales $10^\prime<\theta<100^\prime$, except for II at low redshift which attains values similar to the E-mode around $10^\prime$), but in some cases we find $\ba{M_{\rm B}^2}(\theta)<0$ for II and GG at statistically significant levels. This is unphysical as the aperture mass is given by an integral over a non-negative power spectrum with a non-negative kernel \citep{Schneider98}.

Increasing the minimum angular separation entering the integral of Equation (\ref{eq:map}) to $3^{\prime\prime}$ removes the negative B-modes, at the price that only scales of $10^\prime$ and above are immune to leakage from the E-mode. Even after these modifications, we see a small but significant positive B-mode lensing signal out to large scales. This is unexpected as \citet{hilbert09} showed that multiple light deflections and lens-lens coupling generate B-modes which are less than a per mil of the E-mode signal. Contributions by source redshift clustering \citep{SvWM02} should be small, or zero in the case of the cross-correlations (because the redshift distributions do not overlap).

The limited resolution of the N-body simulation or the shear maps cannot be the cause of $\ba{M_{\rm B}^2}(\theta)<0$ because in that case smaller scales and lower redshifts should be affected more strongly, in contrast to what we observe. A potential explanation is simulation particle shot noise which affects $\xi_+$ only at zero lag but contributes to $\xi_-(\theta)$ at $\theta>0$. Excess signal in $\xi_-$ leads to more negative B-modes than expected from a cosmological signal; see Equation (\ref{eq:map}).

We cannot exclude that the residual GG signal is spurious, and that the II and GI B-mode signals are affected at a similar level. Therefore we defer a detailed investigation of the significance of B-modes generated by intrinsic alignments to future analysis. Note that these systematics affect the E-mode signals at the per cent level and are hence negligible in the preceding analysis.

\section{Conclusions}
\label{sec:conclusions}

In this work we investigated the correlations induced by the intrinsic alignment of galaxies using semi-analytical models of galaxy shapes. These models are based on the dark matter halo properties extracted from the Millennium Simulation,  supplemented by semi-analytic models of galaxy formation and information on the link between dark matter and the luminous part of galaxies from various simulations with and without baryons.

In a first step the resulting intrinsic ellipticity correlations were analysed for central galaxies whose halo properties were measured in the simulation, differentiating between early and late types as identified by the semi-analytic models. Late-type galaxies are only weakly aligned, with small positive correlations on scales below $5\,{\rm Mpc}/h$ which increase marginally with redshift but do not show any strong scaling with mass, luminosity, or environment, although statistical errors remain large for this sample.

Early-type galaxies are strongly aligned with a significant signal beyond physical separations of $r = 20\,{\rm Mpc}/h$ which increases towards smaller separations (three orders of magnitude over two decades in $r$) and towards higher redshifts (one order of magnitude from $z=0$ to $z=2$). We found that the alignment strength increases with halo mass up to $m_{\rm halo} \approx 4 \times 10^{13}\,M_\odot$ beyond which it remains constant, while it does not depend on luminosity or galaxy number density in the environment if the halo mass remains fixed.

We compared our galaxy shape models to a suite of recent observations of intrinsic alignments in both blue- and red-selected galaxy samples. Taking into account a realistic composition of the galaxy sample \citep[see also][]{heymans06} and the effects of satellite galaxies, we obtained combinations of simple shape models that are in good agreement with early-type observations. Models that are based on reduced inertia tensor measurements of dark matter halo shapes (which tend to be closer to spherical), and that include random misalignments of satellite galaxies \citep[taken from][]{knebe08}, yield better fits. A model including random misalignments of central early-type galaxies as proposed by \citet{okumura08,okumura09} is also in fair agreement with observations, but such a misalignment is not required to match the signals from observations and simulations.

Reduced inertia tensor measurements arguably yield a better approximation to the shape of a galaxy as they put more weight towards the centre of a halo. Ideally one would obtain the shape estimate only from the inner parts of the halo, but this increases the requirements on mass resolution as several hundred simulation particles now have to be within the chosen sub-region alone. Dark matter haloes become more spherical towards their centre \citep{schneiderm11}, a trend that becomes more pronounced if the impact of baryons is taken into account \citep{bryan13}. We demonstrated that, depending on how the halo shape is measured, the prediction of intrinsic alignment amplitudes can vary by factors of a few.

While the comparison with the measurement of late-type galaxy alignments by \citet{mandelbaum11} remains inconclusive due to the large statistical uncertainty of the observational data, good agreement of the radial and redshift dependencies between the observations and model predictions for early types was found. None of our models was capable of reproducing the observed luminosity, largely over-predicting alignments for faint objects. This discrepancy could be due to a failure of the model for satellite alignments or indicate that the luminosity scaling found observationally for LRG-like galaxies does not extend to $M_r>-19$. Both hypotheses can be tested with forthcoming simulations and observations.

Using the galaxy shape model that best reproduced the joint observations of two-point statistics and the shape distributions analysed in Paper I, we predicted the intrinsic alignment contamination of a \textit{Euclid}-like survey, measured in terms of the ratio of total E-mode intrinsic alignment (II+GI) over lensing signal, $R_{\rm IA}$. We obtained $R_{\rm IA}$ below $10\,\%$ for $z>0.6$ and on the angular scales relevant for weak lensing analyses, i.e. above a few arcminutes, with higher contamination on smaller scales and at lower redshift.

These results are of the same order as earlier estimates \citep{heymans06} and confirm that intrinsic alignments will be one of the major systematics for forthcoming cosmic shear surveys, but not significantly degrade constraints on cosmology if moderate priors on the intrinsic alignment signal can be imposed (\citealp{joachimi10}; see also \citealp{kirk12} for the impact on cosmological parameter estimation by several intrinsic alignment models).

We investigated the effect of the galaxy sample selection on the strength of intrinsic alignments, selecting survey parameters similar to KiDS and \textit{Euclid}, and applying cuts in observed colour to remove the reddest galaxies. Variations in limiting magnitude and the filter in which the selection is made can change alignment signals, in particular the II term, by factors of a few. The GI signal can almost be halved by reducing the sample size in the foreground by $20\,\%$, corresponding to an increase in statistical error on the lensing signal by less than $12\,\%$. Of course, one can undertake a much more sophisticated sample selection, e.g. by using cuts in rest-frame magnitude and colour, which can be approximately calculated by means of photometric redshifts. This should be more effective at selecting galaxies with particularly low (or high) $R_{\rm IA}$ and will be investigated in future work.

Note that here we only suggested to make the sample selection in different filters. When also performing the shape measurement at different wavelengths, different structures within the galaxies dominate the light distribution, and thus the alignment properties may change as well. We cannot investigate this effect with our simple analytic models, but the forthcoming multi-band surveys will shed more light on its significance. Ultimately, the intrinsic alignment contamination may become a major reason to perform weak lensing studies in a completely different part of the electromagnetic spectrum \citep[see e.g.][]{patel10}.

B-modes generated by intrinsic ellipticity correlations seem to be affected by spurious contributions, so that we can only set an upper limit of $|\ba{M_{\rm B}^2}(\theta)|<10^{-7}$. A potential exception is the II B-mode signal at $z \sim 0.5$, which attains similar amplitudes as the corresponding E-mode. Such signals could serve as an independent complement to E-modes to constrain intrinsic alignment models. Future analysis of highly accurate simulations with very good mass and spatial resolution is needed to firmly establish whether intrinsic alignments leave a clear signature in the B-mode, especially at high redshift.

A semi-analytic approach to galaxy shape modelling as proposed here, or alternatively an advanced halo model \citep[see][]{schneiderm09,schneiderm11}, has the major advantage of jointly predicting various shape and clustering statistics for a wide range of galaxy samples. Such models readily combine constraints from large and small scales and include realistic galaxy bias, allowing access to the deeply non-linear regime and thus enabling two-point analyses of small-scale surveys such as COSMOS.

Although the shape models employed in this work are still simplistic and involve strong assumptions, the actual major concern is the lack of observational constraints. To date, there are essentially no informative intrinsic alignment measurements for blue, under-luminous galaxies which, however, are the main constituents of galaxy samples used for weak lensing. Furthermore the amplitude of intrinsic alignments at redshifts of order unity and beyond is currently unknown although the majority of galaxies in forthcoming weak lensing surveys will populate this regime.

Future observations need to improve statistical constraints and extend them closer to the depths of weak lensing studies. This is a challenging task as both accurate galaxy shape measurements and redshift estimates have to be available, as e.g. in the overlap regions of imaging and spectroscopic redshift surveys. Reducing shot noise and sample variance sets tight requirements on number density, area coverage, and survey geometry for the galaxy sample to be studied. Since a critical ingredient is a redshift estimate for every galaxy which is accurate to roughly the per-cent level, multi-band photometric surveys like PAU\footnote{\texttt{http://www.pausurvey.org/}}, or low-resolution spectroscopy as performed by PRIMUS \citep{coil11}, targeted at weak lensing surveys, are a promising strategy.

The \textit{Euclid} mission not only sets the highest requirements on the control of intrinsic alignments, but also provides data ideally suited for their calibration, with unprecedented constraints expected on emission-line galaxies from the spectroscopic survey, and the opportunity of directly measuring the intrinsic alignment signal on the weak lensing sample thanks to the highly accurate photometric redshifts supplied by the deep survey \citep{laureijs11}.

Next steps in the simulation-based modelling of galaxy shapes include the usage of a hierarchical suite of N-body simulations which will provide the cosmological volume to improve on statistical power, as well as the mass resolution to determine robust halo properties of low-mass and satellite galaxies. Semi-analytical models that simultaneously determine the photometry and the morphology, based on halo properties and merger history, will establish a more realistic link between the shape of a galaxy and its luminosity and colours, thereby allowing us to examine the efficiency of suppressing the contamination by intrinsic alignments via colour cuts on the galaxy sample.

The results of the hydrodynamic simulations performed by \citet{sales12} challenge the paradigm that the different evolution of disc and spheroidal galaxies is caused by the details of the merger history. They found that the difference between late and early galaxy types can rather be associated with the infall directions and thermo-dynamic state of the accreted gas. Similarly, \citet{kimm12} concluded from their suite of hydrodynamic simulations that the angular momentum of galactic gas is larger than that of the dark matter halo, and closely linked to the surrounding large-scale structure.

These recent findings question the choice of analytic models of galaxy shapes that are directly derived from halo properties, as taken in this paper. However, the proposed scenarios are likely to produce distinctively different predictions for e.g. the radial dependence (due to a stronger link of angular momentum with filamentary structures) and the evolution (due to a stronger correlation with gas infall events than with mergers) of intrinsic alignments, and hence can be put to the test by comparison to intrinsic alignment observations in analogy to our approach.

Once hydro-dynamic simulations can be run on cosmologically relevant scales with sufficient resolution, one can by-pass the critical assumptions about the link between the shapes of dark and luminous matter and directly create mock galaxy images on which a standard shape measurement code can then be run. There would still be critical assumptions involved, most prominently about the sub-grid physics of baryons in the simulation. For instance, a strong dependence on feedback models, as seen for weak lensing power spectra \citep{semboloni11b}, is also expected for intrinsic alignment signals. However, these issues would be shared with the models of galaxy evolution and of cosmological probes, and hence allow for intrinsic alignments to be integrated smoothly into a wider cosmological context.

\section*{Acknowledgments}

We are grateful to Rachel Mandelbaum for providing us with the WiggleZ correlation function data, and to Catherine Heymans for providing us with code to compute aperture mass statistics. Moreover we would like to thank them, as well as Chris Blake, Tim Schrabback, Andy Taylor, and Ismael Tereno for helpful discussions. We are grateful to our referee for constructive remarks.

BJ acknowledges support by an STFC Ernest Rutherford Fellowship, grant reference ST/J004421/1. ES and HH acknowledge the support of the Netherlands Organization for Scientific Research (NWO) through a VIDI grant, as well as support from the ERC under FP7 grant number 279396. PEB acknowledges support by the Deutsche Forschungsgemeinschaft (DFG) under the project SCHN 342/7–1 in the framework of the Priority Programme SPP-1177, and the Initiative and Networking Fund of the Helmholtz Association, contract HA-101 (‘Physics at the Terascale’). JH, SH, and PS acknowledge support by the DFG through the Priority Programme 1177 `Galaxy Evolution' (SCHN 342/6 and WH 6/3) and the Transregional Collaborative Research Centre TRR 33 `The Dark Universe'. SH also acknowledges support by the National Science Foundation (NSF) grant number AST-0807458-002.

The simulations used in this paper were carried out as part of the programme of the Virgo Consortium on the Regatta supercomputer of the Computing Centre of the Max-Planck-Society in Garching, and the Cosmology Machine supercomputer at the Institute for Computational Cosmology, Durham. The Cosmology Machine is part of the DiRAC Facility jointly funded by STFC, the Large Facilities Capital Fund of BIS, and Durham University.

\bibliographystyle{mn2e}

\label{lastpage}
\end{document}